%% file: ms.tex
\documentclass[preprint]{aastex}
\usepackage{natbib}
\shorttitle{Radiating Current Sheets}
\shortauthors{Goodman and Judge} 

\begin{document}

\title{Radiating Current Sheets in the Solar Chromosphere} 
\author{Michael L. Goodman}        
\affil{Advanced Technologies Group, West Virginia High Technology Consortium Foundation\\1000 Galliher Drive, Fairmont, WV 26554}
\email{mgoodman@wvhtf.org}

\and 

\author{Philip G. Judge} \affil{High Altitude Observatory, National
Center for Atmospheric Research\altaffilmark{1}, P.O. Box 3000,
Boulder CO~80307-3000, USA\\ \vbox{}}
\email{judge@ucar.edu}

\altaffiltext{1}{The National Center for Atmospheric Research is
sponsored by the National Science Foundation}

\begin{abstract}
An MHD model of a Hydrogen plasma with flow, an energy equation, NLTE ionization and radiative cooling, and an Ohm's law with anisotropic electrical conduction and thermoelectric effects is used to self-consistently generate atmospheric
layers over a $50$ km height range. A subset of these solutions contain current sheets, and have properties similar to those of the lower and middle chromosphere. The magnetic field profiles are found to be close to Harris sheet profiles, with maximum field strengths $\sim 25-150$ G. The radiative flux $F_R$ emitted by individual sheets is $\sim 4.9 \times 10^5 - 4.5 \times 10^6$ ergs-cm$^{-2}$-s$^{-1}$, to be compared with the observed chromospheric emission rate of $\sim 10^7$ ergs-cm$^{-2}$-s$^{-1}$. Essentially all emission is from regions with thicknesses $\sim 0.5 - 13$ km containing the neutral sheet. About half of $F_R$ comes from sub-regions with thicknesses 10 times smaller. A resolution $\lesssim 5-130$ m is needed to resolve the properties of the sheets. The sheets have total H densities $\sim 10^{13}-10^{15}$ cm$^{-3}$. The ionization fraction in the sheets is $\sim 2-20$ times larger, and the temperature is $\sim 2000-3000$ K higher than in the surrounding plasma. The Joule heating flux $F_J$ exceeds $F_R$ by $\sim 4-34 \%$, the difference being balanced in the energy equation mainly by a negative compressive heating flux. Proton Pedersen current dissipation generates $\sim 62-77\%$ of the positive contribution to $F_J$. The remainder of this contribution is due to electron current dissipation near the neutral sheet where the plasma is weakly magnetized. 
\end{abstract}

\section*{1. Introduction}

Understanding diffusive transport processes in the chromosphere is the key to understanding how it is heated, and how it couples to the corona. These processes include resistive and viscous dissipation, thermal conduction, and thermoelectric effects. They are anisotropic due to the strong magnetization of the chromosphere. Since transport processes play a major role in determining the plasma temperature $T$, they play a major role in determining the radiative emission, which is dominated by inelastic collisions of atomic ions with electrons. Models that include accurate representations of these processes, and that are solved with sufficient resolution to resolve the length scales associated with these processes are necessary for accurately computing their effects.

One such model is presented here. It is remarkable that the model self-consistently generates solutions containing
current sheets under solar chromospheric conditions, and that radiative emission from these sheets is large enough so
that the collective emission from a reasonable number of them can account for the observed chromospheric radiative loss of $\sim 10^7$ ergs-cm$^{-2}$-sec$^{-1}$. The Joule heating that drives this emission does not involve Sweet-Parker type magnetic reconnection (Parker 1979, 1994; Biskamp 1997, 2000, 2003; Priest \& Forbes 2000; Boozer 2004, 2005; Yamada 2007; Yamada, Kulsrud \& Ji 2010) since the flow velocity is normal to the sheets, and cannot change sign due to mass conservation. It may be that such current sheets are sites of significant chromospheric heating before or without undergoing reconnection. 

It is equally remarkable that these current sheet solutions arise spontaneously as a result of the intrinsic transport
physics of the model. There is no a priori specification of solutions to the model to cause it to generate solutions of this or any other type. 

The current sheet solutions might be relevant to magnetic flux emergence, and associated heating and emission in the real
Sun. Since the model is 1D with variation along the direction of gravity, and since the vertical magnetic field is
assumed zero, any current sheets generated by the model must be horizontal current sheets. Solar observations indicate that magnetic bipoles on the scale of the granulation continually emerge in quiet and active regions, and that some of
them rise into the chromosphere while their footpoint separations increase to $\sim 500 - 4000$ km, creating an extended
horizontal magnetic field between the footpoints that may form horizontal current sheets as it rises into regions with
opposite horizontal magnetic polarity (Guglielmino, Zuccarello, Romano \& Bellot Rubio 2008; Mart\'{i}nez Gonz\'{a}lez \& Bellot Rubio 2009; Lites 2009; Petrie \& Patrikeeva 2009). These bipoles have lifetimes $\sim 2- 40$ minutes, and field strengths up to several hectoGauss (Mart\'{i}nez  Gonz\'{a}lez \& Bellot Rubio 2009). Enhanced emission with a duration $\sim 30$ minutes is observed to be associated with bipoles rising into the chromosphere, and is suggested as being due to the formation, and heating within current sheets formed as mentioned above (Guglielmino, Zuccarello, Romano \& Bellot Rubio 2008).
This process, along with current sheet formation that might occur between vertical sections of bipoles with opposite
magnetic polarity (Shibata et al. 2007; Chae et al. 2010), might be a significant source of chromospheric heating.
The current sheet solutions presented here show that magnetic reconnection is not the only possible explanation of thermal chromospheric phenomena involving flux emergence.

\section*{2. The Model}

All quantities are assumed to depend only on height $z$ in Cartesian coordinates $(x,y,z)$. A steady state is assumed,
and it is assumed that $B_y=B_z=0$, where ${\bf B}$ is the magnetic field. Let ${\bf V}$ be the center of mass (CM) velocity. Omitting viscous effects, the $x$ and $y$ components of the momentum equation reduce to $V_x$ and $V_y$ being
arbitrary constants. $V_x$ does not play any role in determining the properties of the solutions, and $V_y$
enters the model only through the vertical electric field $E_z$, which does not play an important role in determining
the solutions.\footnote{The values of $V_x$ and $V_y$ might be important in a model that uses the current sheet solutions presented here as initial states, and determines their evolution in time. The simplest example is a linear model for determining the stability of the current sheets, and the linear waves they support.}
The transport processes included in the model are anisotropic electrical conduction, thermoelectric current drive, and NLTE radiative cooling. They are respectively represented in the model by an electrical conductivity tensor and thermoelectric term in the Ohm's law, and by a radiative cooling term in the energy equation. 

The mass conservation equation reduces to $\rho V_z=M$, where $M$ is the constant vertical mass flux. The $z$ component of the momentum equation is 
\begin{equation}
(M V_z + p + B_x^2/8\pi)^\prime = - \rho g. \label{mom1} 
\end{equation}
Here $p,\rho$, and $g$ are the
total pressure, total mass density, and gravitational acceleration, and the prime denotes $d/dz$. 
The electrostatic condition $\nabla \times {\bf E}=0$ reduces to $E_x$ and $E_y$ being constant. 

Since a pure H atmosphere is assumed along with quasi-neutrality and the ideal gas equation of state for each species, the electron, proton, and HI number densities $n_e, n_p$, and $n_H$ may be written in terms of $p,\rho$, and $T$ as 
\begin{eqnarray}
n_e&=&\frac{p}{k_B T} - \frac{\rho}{m_p} = n_p \label{ne}\\
n_H&=&\frac{2 \rho}{m_p} - \frac{p}{k_B T}.\label{nH}
\end{eqnarray}
Here $k_B$ is Boltzmann's constant.

\subsection*{2.1. Ohm's Law}

The Ohm's law is a reduced form of the exact generalized Ohm's law for a three species plasma derived by Mitchner 
\& Kruger (1973). This reduced Ohm's law results from imposing long time-scale approximations on the system of particle momentum equations, neglecting viscous effects, and assuming the net ionization rate of HI is zero. Under typical chromospheric conditions, the long time scale approximations translate into the constraint that the characteristic minimum time scale $t_c$ over which a significant change in the macroscopic state of the plasma occurs satisfies $t_c^{-1} \ll$ a few hundred Hz\footnote{The long time scale approximations are equivalent to neglecting the displacement current, and imposing the constraints $(\nu_{ep}+ \nu_{eH}) t_c \gg 1$, and $(m_p/m_e)(\nu_{pH}+(m_e/m_p)^{1/2} \nu_{pe}) t_c \gg 1$. These constraints allow certain viscous terms to be neglected.
Here $\nu_{eH},\nu_{ep},\nu_{pe}$, and $\nu_{pH}$ are the momentum transfer collision frequencies for e-HI, e-p, p-e, and p-HI collisions.}.
This constraint is satisfied for the current sheets generated here. The Ohm's law used here is
\begin{equation}  
{\bf E}_{CM} = -\frac{\nabla p_e}{e n_e} +\eta \left\{{\bf J} +M_e {\bf J} \times \hat{\bf B} + \Gamma \hat{\bf B} \times \left( {\bf J} \times \hat{\bf B} \right) - \frac{\Gamma c}{B \xi_H} \hat{\bf B} \times \left( \xi_H \nabla p - 
\nabla p_H  \right) \right\}. \label{ohm}
\end{equation}
Here ${\bf E}_{CM} \equiv {\bf E} + ({\bf V} \times {\bf B})/c$ is the center of mass electric field, ${\bf E}$ is
the electric field, $\xi_H = \rho_H/\rho$ is the ratio of neutral to total mass densities, $\hat{\bf B}$ is a unit vector in the direction of ${\bf B}$, $p_H=n_H k_B T$ is the HI pressure, $\eta$ is the parallel (Spitzer) resistivity, and 
$\Gamma = \xi_H^2 M_e M_p$, where $M_e$ and $M_p$ are the electron and proton magnetizations. The magnetization of a
particle species $s$ is $M_s = \omega_s/\nu_s$, where $\nu_s$ and $\omega_s$ are the total momentum transfer collision
frequency and the cyclotron frequency of species $s$.

The term $(\Gamma c/B \xi_H) \hat{\bf B} \times (\xi_H \nabla p - \nabla p_H)$ in equation (\ref{ohm}) is sometimes neglected under the assumption that $\xi_H \sim 1$, and nearly uniform. If $\xi_H \sim 1$, this term reduces to $(2 \Gamma c/B) \hat{\bf B} \times \nabla p_e$, since $p= p_H + 2p_e$. Comparing this term with the Hall term $M_e {\bf J} \times \hat{\bf B}$ in the Ohm's law, and assuming that $({\bf J} \times {\bf B})/c \sim \nabla p$ based on force balance, it follows that $(2 \Gamma c/B) \hat{\bf B} \times \nabla p_e$ may be neglected relative to the Hall term if the condition $|2 M_p \nabla_\perp p_e| \ll |\nabla_\perp p|$ is satisfied. Here $\nabla_\perp$ denotes the gradient $\perp {\bf B}$. Since $p_e \rightarrow 0$ as $\xi_H \rightarrow 1$, it follows that this condition is satisfied for sufficiently low ionization since $M_p$ is bounded. No assumptions are made about $\xi_H$ in this paper. The full effect of the term $(\Gamma c/B \xi_H) \hat{\bf B} \times (\xi_H \nabla p - \nabla p_H)$ is included. It is found to have a small to moderate
thermoelectric effect that reduces the Joule heating rate in the current sheet solutions presented in \S 4.

For the model considered here the $x,y$, and $z$ components of the Ohm's law reduce to $E_x=0$,
\begin{eqnarray}
E_y + \frac{V_z B_x}{c} &=& \frac{c \eta}{4 \pi} \left( (1+ \Gamma) B_x^\prime + \frac{4 \pi \Gamma}{B_x \xi_H} 
(\xi_H p^\prime - p_H^\prime)\right), \label{ohmy1}\\
E_z -\frac{V_y B_x}{c} &=& -\frac{p_e^\prime}{e n_e} - \frac{\eta c M_e B_x B_x^\prime}{4 \pi |B_x|} \label{ohmz}.
\end{eqnarray}

\subsection*{2.2. NLTE Saha Equation}

For a hydrogen plasma out of LTE, the ionization balance (equivalent of the LTE Saha equation) determining $n_H, n_p$, 
and $n_e(=n_p)$ can be approximated from earlier work (e.g. Hartmann \& MacGregor 1980) by considering a simplified hydrogen atom sitting above the solar photosphere in chromospheric plasma.  To a first approximation, we can assume radiative detailed balance in the Lyman lines and continuum, as the bulk of the chromosphere is effectively thick in these transitions.  Ionization of hydrogen occurs
through collisions with electrons, but has a strong contribution from the flux of photospheric light in the Balmer continuum.  Recombination effectively occurs to the $n=2$ levels ($n=$ the principal quantum number), and higher because
the Lyman continuum is in detailed balance (photoionization balances radiative recombination). Under these conditions, photoionization in the Balmer continuum, in cm$^{-3}$-s$^{-1}$, occurs at a rate of 8000 s$^{-1}$ for each atom in the $n=2$ quantum levels (Vernazza, Avrett \& Loeser 1981), or 
\begin{equation}
R_i= 3.2 \times 10^4 n_{H} \exp(-T_2/T). \label{rion}
\end{equation}
Here $T_2= e E_{n2}/k_B$, where $E_{n2} = (3/4) \times 13.58$ eV is the $n=2$ level energy, and  $T_2 \sim 1.18187 \times 10^5$ K.

The radiative recombination rate in the same units is
\begin{equation}
R_r = \alpha n_e n_p \left(\frac{T}{10^4}\right)^b \label{rrecomb}.
\end{equation}
We have computed recombination only to the $n=2$ levels and higher which yields 
$\alpha = 3.373 \times 10^{-13}$, and $b = -0.786$.

Assuming statistical equilibrium implies $R_i = R_r$. This equality yields the following equation for $n_e$ as a function of $n_H$ and $T$, noting that $n_e = n_p$.
\begin{eqnarray}
n_e &=& n_H^{1/2} \left(\frac{3.2 \times 10^4}{\alpha}\right)^{1/2} \left(\frac{T}{10^4}\right)^{|b|/2} 
\exp\left(-T_2/2T\right).  \label{saha} \\
&\equiv& n_H^{1/2} I(T) \label{saha1}
\end{eqnarray}
Equation (\ref{saha}) includes all lowest order NLTE effects under solar chromospheric conditions. 

Combining equation (\ref{saha1}) with equations (\ref{ne}) and (\ref{nH}) for $n_e(p,T)$, and $n_H(p,t)$ gives the following equation for $\rho(p,T)$, which is used in the equations in \S 3 that are
solved numerically to determine the state of the plasma. 
\begin{equation}
\rho(p,T) = m_p\left(I^2(T) + \frac{p}{k_B T}-I(T) \left(I^2(T)+ \frac{p}{k_B T}\right)^{1/2}\right). \label{rho}
\end{equation}

\subsection*{2.3 Effectively Thin Radiative Loss Rate}

Radiation losses represent a dominant sink of energy from chromospheric plasma.  Decades of radiative transport
calculations enable us to treat, at a similar level of approximation to the ionization balance, these losses as if they are 
``effectively thin'':  the chromosphere has strong contributions from trace elements in spite of their lower abundances, 
because lines of H and He are effectively thick in the deeper chromospheric layers and also they require enormous
energies to excite them ($T_2 \gg T$, for example). Anderson \& Athay (1989) found that an effectively thin approximation works 
to within a factor of two in comparison with detailed transfer calculations, because lines of Fe~II, Ca~II, and Mg~II
separately tend to dominate the radiation losses each under effectively thin conditions at the different heights where they are formed. 
Therefore, we have computed the radiation losses under effectively thin conditions, assuming again the Lyman lines
contribute nothing to the losses, using data from the CHIANTI database (see the Appendix).  Below $T=1.5\times 10^4$K, 
the effectively thin radiative loss rate, in ergs-cm$^{-3}$-sec$^{-1}$, is represented by 
\begin{eqnarray}
Q_R &=& 8.63 \times 10^{-6} C_E \frac{n_e (n_{H} + n_p)}{T^{1/2}} \Sigma_{i=1}^2 E_i \Upsilon_i \exp(-e E_i/k_B T). \label{Qrad} \\
&\equiv& R(T) n_e(n_H + n_p) \label{Qrad1}
\end{eqnarray}  
The form of this solution is a physical fit to a three level generic ``hydrogen atom'' with two excited states at energies 
$E_{1,2}$ in eV, excited by collisions from a ground state.  The quantities 
$\Upsilon_i$ are in the form of Maxwellian-averaged collision strengths. To match the 
radiation losses from all the elements their values were found to be 
$E_1=3.54$ eV, $E_2 = 8.28$ eV, $\Upsilon_1 = 0.15 \times 10^{-3}$, and $\Upsilon_2 = 0.065$. The temperature $T$ is in K, 
$C_E=1.6022 \times 10^{-12}$ ergs-eV$^{-1}$, and $n_e, n_p$, and $n_{H}$ are in cm$^{-3}$.

As shown in the Appendix, below $1.5\times10^4$ K, this three level mean H atom radiates the same energy as a solar plasma with the solar photospheric trace element abundances. Equation (\ref{Qrad}) for $Q_R$ is accurate to within $\sim 10 \%$ for $T \leq 1.5 \times 10^4$ K.

\subsection*{2.4. Thermal Energy Equation}

The thermal and ionization energy per unit volume is $3 p/2 + \chi_H n_e$, where $\chi_H$ is the ionization energy of 
H\footnote{Note that 1/4 of the energy used to ionize hydrogen is provided by incident photospheric radiation external to the model, 
and is not supplied by local electron thermal energy.}.
The thermal energy equation is
\begin{equation}
\left(\left(\frac{3 p}{2} + \chi_H n_e\right) V_z\right)^\prime = - p V_z^\prime + \frac{c}{4 \pi} B_x^\prime \left(E_y + \frac{V_z B_x}{c}\right) - Q_R. \label{energy}
\end{equation}
The terms on the right hand side (RHS) of this equation are the compressive and Joule heating rates, and the radiative cooling rate. The first two rates can be positive or negative. Joule dissipation and compressive heating are the local thermal energy generation mechanisms included in the model, respectively converting electromagnetic and CM kinetic energy into thermal energy and vice versa. Viscous heating and diffusive thermal energy flux are omitted.

\section*{3. Equations for Numerical Solution}

Using $\rho V_z=M$, and equations (\ref{ne}), (\ref{nH}), and (\ref{rho}), the momentum, Ohm's law, and energy equations 
(\ref{mom1}), (\ref{ohmy1}), and (\ref{energy}) may be written as a set of three coupled equations involving only $p,T$,
and $B_x$.

The momentum equation (\ref{mom1}) becomes
\begin{eqnarray}
&&\left[1- \frac{M^2 m_p}{2 k_B T \rho^2} \left(3-\frac{m_p p}{\rho k_B T}\right)\right] p^\prime + \frac{M^2 m_p}{2 T \rho^2} \left[\left(\frac{\rho}{m_p}\right)^{1/2} \left(2 - 
\frac{m_p p}{\rho k_B T}\right)^{3/2} \times  \right. \nonumber \\ 
&&\left. I \left(|b|+ \frac{T_2}{T}\right) +\frac{p}{k_B T} \left(3-\frac{m_p p}{\rho k_B T} \right)\right] T^\prime + \frac{B_x B_x^\prime}{4 \pi} = - \rho g. \label{momnum}
\end{eqnarray}

The Ohm's law equation (\ref{ohmy1}) becomes
\begin{equation}
E_y+\frac{M B_x}{\rho c} = \eta \left\{\frac{c}{4 \pi} \left(1+ \Gamma\right) B_x^\prime + \frac{\Gamma c}{B_x} \left(\frac{p}{k_B T} - \frac{\rho}{m_p} \right) \left(1 + |b| + \frac{T_2}{T}\right) k_B T^\prime \right\}. \label{ohmnum}
\end{equation}
The parallel resistivity $\eta$, and the magnetization induced resistivity $\eta \Gamma$ are expressed as functions of
$p,T$, and $B_x$ as follows
\begin{eqnarray}
\eta &=& \frac{m_e(\nu_{ep}+\nu_{eH})}{n_e e^2} \\
&=& m_e^{1/2} \left\{\frac{4 (2 \pi)^{1/2} e^2 \ln \Lambda}{3 (k_B T)^{3/2}}+ \frac{\sigma (2 \rho k_B T - m_p p)
(k_B T)^{1/2}}{e^2 (m_p p - \rho k_B T)} \right\} \label{eta} \\
\Lambda &=& \frac{3 m_p^{1/2}(k_B T)^2}{2 e^3 \pi^{1/2}(m_p p - \rho k_B T)^{1/2}} \\
\eta \Gamma &=&\frac{(2 \rho k_B T - m_p p) B^2 m_p^{3/2}}{(m_p p - \rho k_B T) \rho^2 c^2 \sigma (k_B T)^{1/2}}. 
\end{eqnarray}
Here $\ln \Lambda$ is the Coulomb logarithm, and the electron-HI scattering cross section $\sigma = 5 \times 10^{-15}$ 
cm$^2$ (Osterbrock 1961 \S VI; Krall \& Trivelpiece 1986 \S 6.8.3). The magnetization factor $\Gamma \propto ((\nu_{ep}+\nu_{eH}) \nu_{pH})^{-1}$, where $\nu_{p H} = n_H \sigma (k_B T/m_p)^{1/2}$.

The thermal energy equation becomes
\begin{eqnarray}
&&\frac{M}{2 \rho} \left\{ 3-\frac{5 p m_p}{2 \rho k_B T} \left(3- \frac{m_p p}{\rho k_B T}\right) + \frac{\chi_H}{k_B T} \left(\frac{m_p p}{\rho k_B T} - 2 \right) \left(\frac{m_p p}{\rho k_B T} - 1
\right) \right\} p^\prime  + \nonumber \\
&&\frac{M p}{4 \rho k_B T} \left\{\frac{5 m_p}{\rho} \left[\left(\frac{\rho}{m_p}\right)^{1/2} \left(2 - \frac{m_p p}{\rho k_B T} \right)^{3/2} I\left(|b|+\frac{T_2}{T}\right) + 
\frac{p}{k_B T} \times \right. \right. \nonumber \\
&&\left. \left. \left(3-\frac{m_p p}{\rho k_B T} \right) \right] + \frac{2 \chi_H}{k_B T} \left(2 - \frac{m_p p}{\rho k_B T} \right) \left[ \left(\frac{m_p}{\rho}\right)^{1/2}  \left(2 - \frac{m_p p}{\rho k_B T} \right)^{1/2}  I\left(|b|+\frac{T_2}{T}\right) +  \right. \right. \nonumber \\
&& \left. \left. \left(\frac{m_p p}{\rho k_B T} - 1\right) \right] \right\} k_B T^\prime = \frac{c}{4 \pi} B_x^\prime \left(E_y + \frac{M B_x}{\rho c}\right) - \frac{\rho}{m_p} \left(\frac{p}{k_B T} - \frac{\rho}{m_p}\right)  R(T). \label{energynum}
\end{eqnarray}

In equations (\ref{momnum})-(\ref{energynum}), $\rho(p,T)$ is given by equation (\ref{rho}). The solution of equations (\ref{momnum}), (\ref{ohmnum}), and (\ref{energynum}) for $T,p$, and $B_x$ is uniquely determined by specifying the values of the constant parameters $E_y$ and $M$, and specifying the boundary conditions $T(0),p(0)$, and $B_x(0)$ at an initial height denoted as $z=0$. The equations are then integrated along the direction of increasing height. 

\subsection*{3.1. Joule Heating Rate}

The Joule heating rate $Q_J$ and its components are obtained by multiplying equation (\ref{ohmnum}) by $J_y$. The equation
becomes $Q_J=Q_S+Q_P+Q_T$, where $Q_S,Q_P$, and $Q_T$ are the Spitzer, Pedersen, and thermoelectric heating rates defined below\footnote{These quantities give the rate of dissipation of energy in the ordered motion in ${\bf J}$ by
collisions between hydrogen, protons, and electrons. ${\bf J}$ is $\propto$ the relative velocity of the ion and electron fluids, and so represents a directed flux of fluid kinetic energy available for conversion into thermal energy.}. 

The first term on the RHS of equation (\ref{ohmnum}) gives rise to a resistive heating rate $Q_S+Q_P \equiv (\eta + 
\Gamma \eta)J_y^2 > 0$, where $Q_P \propto$ the magnetization induced resistivity $\Gamma \eta$. For the current sheet solutions presented in \S 4, $Q_P$ is the main driver of the radiative cooling rate.

The second term on the RHS side of equation (\ref{ohmnum}) is the term $(\Gamma c/B \xi_H) \hat{\bf B} \times (\xi_H \nabla p - \nabla p_H)$ in equation (\ref{ohm}) in \S 1.1. This term corresponds to a thermoelectric contribution $Q_T$
to $Q$, since it is $\propto T^\prime$, given by
\begin{equation}
Q_T = \frac{\Gamma c^2 \eta B_x^\prime}{4 \pi B_x} \left(\frac{p}{k_B T} - \frac{\rho}{m_p} \right) \left(1 + |b| + \frac{T_2}{T}\right) k_B T^\prime.
\end{equation}
Insight into the sign of $Q_T$ is obtained as follows. Define $\bar{\Gamma}$ by $\Gamma = \bar{\Gamma} B_x^2$. Then 
$\bar{\Gamma} > 0$, and since $p/(k_B T) - \rho/m_p= n_e$, $Q_T$ may be written as
\begin{equation}
Q_T = \frac{\bar{\Gamma} c^2 \eta n_e}{8 \pi}(B_x^2)^\prime \left(1 + |b| + \frac{T_2}{T}\right) k_B T^\prime.
\end{equation}
The sign($Q_T$)=sign($(B_x^2)^\prime T^\prime$), so $Q_T$ may be positive or negative.

\section*{4. Numerical Solutions}

Four solutions are presented in four ranges of density, temperature, and magnetic field strength believed to represent 
conditions in the lower to the mid-chromosphere. This is the region of the chromosphere that emits almost all of the net
chromospheric radiative loss of $\sim 10^7$ ergs-cm$^{-2}$-s$^{-1}$. For the solutions presented here, the largest
radiative loss rate is from current sheets in the lower chromosphere, and the radiative loss rate tends to decrease with
increasing height, in qualitative agreement with the losses predicted by semi-empirical models (e.g. Vernazza, Avrett \& Loeser 1981). 
 
A way to derive the values of the model inputs that cause the model to generate current sheets is not known. The inputs used to generate the current sheet solutions presented here are obtained using heuristic guidelines discussed in \S 6.1. 
The boundary conditions and input parameters for the solutions presented here are as follows. 
Define a set of characteristic densities $(n_1,n_2,n_3,n_4)=(4 \times 10^{12},10^{13},10^{14},10^{15})$ cm$^{-3}$. 
Then the four sets of inputs that determine the solutions are
$(B_x(0)(\mbox{G}),T(0)(\mbox{K}),p(0)(\mbox{dynes-cm$^{-2}$}),V_z(0)(\mbox{m-s$^{-1}$}),\\E_y(\mbox{V-m$^{-1}$}))=$ $(25,7500,k_B n_1 7500,46,-0.1182)$,$(50,6500,k_B n_2 6500,30,-0.1951)$, $(50,5000,k_B n_3 5000,$ $4.84,0.1211)$, and $(100,4000,k_B n_3 4000,4,-0.2452)$. Here $V_z(0)$ is specified instead of the vertical mass flux $M$ since given $p(0), T(0)$, and $V_z(0)$, $\rho(0)$ is known from equation (\ref{rho}), and $M=V_z(0)/\rho(0)$. 

Given these initial values, the corresponding total H density at $z=0$ is $n(0)= n_H(0) + n_p(0) =$$(3.8 \times 
10^{12},9.91 \times 10^{12},10^{14},10^{15})$ cm$^{-3}$. These values of $n(0)$ are used to label the solutions in the
figures. The solutions are labeled in order of increasing density, and are proposed to correspond to decreasing height in
the chromosphere.

The solutions are generated over a 50 km height range. For each solution the current sheet is found to occupy a small fraction of this range.  

\subsection*{4.1. Harris Type Current Sheets}

Figure 1 shows $B_x$ (left panel) and $J_y$ (right panel) vs. height for the four solutions. The profiles of $B_x$ are close to Harris sheet profiles. There is nothing deliberately built into the model or its inputs to make $B_x$ have these profiles.

Not all solutions to the model contain current sheets. Smooth solutions for chromospheric atmospheres extending over a
260 km height range that do not contain current sheets are also found, but not presented here. These atmospheres are
weakly heated and weakly radiating compared to the current sheet solutions, and to the energy requirements of the
chromosphere.\footnote{The full range of solutions to the model is not yet known. Since five inputs are needed to specify the solution, and solutions do not exist for many sets of inputs, it is a nontrivial task to determine the full range of solutions.}

%


\subsection*{4.2  Length Scales, Radiative Emission Rate, and Required Numerical Resolution}

From the Ohm's law equation (\ref{ohmy1}) the resistive diffusivity is $c^2(1+ \Gamma)\eta/(4 \pi)$. Dividing this by
$V_z$ gives the resistive length scale
\begin{equation} 
L_r = \frac{c^2(1+ \Gamma)\eta}{4 \pi V_z}. \label{reslengthscale}
\end{equation}
Figure 2 (left panel) shows $L_r$ vs. height for each solution. It is found that $L_r$ characterizes the thickness of the current sheets in that essentially all of the radiative emission is from a region with a thickness equal to a few times the minimum value of $L_r$ for a given sheet. This region is centered around the height at which $Q_R$ is a maximum and 
$L_r$ is a minimum. 

Figure 2 (right panel) shows $Q_R$. The radiative flux $F_R$ is the integral of $Q_R$ over the height range of the solutions. For the four sheets, in order of increasing $n(0)$, $F_R=(0.488,1.55,1.03,4.45) \times 10^6$ ergs-cm$^{-2}$-s$^{-1}$.
Although the maximum peak values of $Q_R$ occur for the smallest values of $n(0)$, the overall trend is for $F_R$ to
increase as $n(0)$ increases. This is due to the increase of $L_r$ as $n(0)$ increases. Although the peak values of $Q_R$ tend to decrease with increasing $n(0)$, $L_r$ increases enough with increasing $n(0)$ to make $F_R$ follow the same trend. The sheets are strongly radiating in that each sheet radiates $\sim 5-45 \%$ of the observed chromospheric radiative loss, with the percentage increasing with increasing $n(0)$, interpreted as decreasing height in the chromosphere.

Magnification of the profiles for $Q_R$ show that essentially all of $F_R$ is due to emission from a region with a thickness $\sim 2-4$ times the minimum value of $L_r$ for a given solution. These regions are centered around the height where $Q_R$ is a maximum. These thicknesses are $\sim 0.5,0.7,2.7$, and $13$ km. It is also found that for the four sheets $\sim 50,50,40$, and $70 \%$ of $F_R$ is due to emission from a sub-region 10 times thinner. 

The sheets are fully resolved since the minimum allowed step size of the adaptive step code used to solve equations 
(\ref{momnum})-(\ref{energynum}) is many orders of magnitude smaller than the smallest values of $L_r$.

These results suggest that simulations and observations need a resolution $\sim 5-100$ m to accurately resolve, and
compute the emission from strongly radiating current sheets in the most strongly radiating regions of the chromosphere. 

\subsection*{4.3. Temperature, Ionization, Density, and Pressure}

Figures 3 and 4 show the temperature, ionization fraction $n_e/(n_e+n_H)$, and electron and HI densities. Figure 5 shows the pressure. The temperature in the sheets is $\sim 2000-3000$ K higher than in the surrounding plasma.
The maximum sheet temperatures are 5800-9000 K. The ionization fraction in the sheets is $\sim 2-20$ times higher than in the surrounding plasma. The maximum values of the ionization fraction are $\sim 2.7 \times 10^{-4}-8.4 \times 10^{-2}$.
The electron density in the sheets is $\sim 10-100$ times higher than in the surrounding plasma. The maximum neutral density in the sheets is $\sim 1.2-3$ times larger than in the surrounding plasma. The pressure experiences a similar
increase since the plasma is mainly neutral, and since the maximum temperature in the sheets is $\sim 1.5$ times higher
than the temperature in the surrounding plasma. 

The results of \S\S 4.1-4.3 show that the current sheet solutions are characterized by plasma conditions believed to exist in the lower and middle chromosphere. As $n(0)$ increases, the conditions within and near the corresponding sheet vary from those characteristic of the middle chromosphere to those characteristic of the lower chromosphere.  

The sheets are so strongly radiating that, for increasing $n(0)$, the number of sheets needed to emit the total net radiative flux from the chromosphere is $\sim 20,7,10$, and 2.  
This result is discussed further in \S 5 in the context of the constraints that observations place upon the existence and filling factor of such strongly radiating sheets.

\subsection*{4.4. Thermal Energy Balance and Driver of the Radiative Emission}

The thermal energy equation (\ref{energy}) may be written symbolically as $Q_R=Q_J + Q_c - Q_{ce}$, which indicates the energy sources that determine the radiative emission. Here $Q_c \equiv -p \nabla \cdot {\bf V} = - p V_z^\prime$ is the
compressive heating rate, and $Q_{ce} \equiv \nabla \cdot \left(\left(3 p/2 + \chi_H n_e\right) {\bf V}\right)= \left(\left(3 p/2 + \chi_H n_e\right) V_z\right)^\prime$ is the heating rate due to the convective energy (ce) flux of thermal and ionization energy, which describes the
net flow of this energy into a region. 

Figures 6 and 7 show $Q_J, Q_R$, and $Q_c-Q_{ce}$ for each of the solutions. $Q_c-Q_{ce}$ does not make a significant contribution to $F_R$, but plays an
important role in energy balance outside the sheets where $Q_R$ is not significant.

Define the Joule, compressive, and convective energy fluxes $F_J, F_c$, and $F_{ce}$ as the integrals of $Q_J, Q_c$, and 
$Q_{ce}$ over the height range of a solution. Then $F_R=F_J + F_c - F_{ce}$. 

In order of decreasing $n(0)$ the values of these fluxes for each of the four sheets, in units of $10^6$ 
ergs-cm$^{-2}$-s$^{-1}$, are $(F_R,F_J,F_c,F_{ce})=$$(4.4507,5.9742,-1.3915,-0.13206)$, $(1.0253,1.0845,-0.0561,$ $-0.00311)$, 
$(1.5499,1.6035,-0.0444,-0.00914)$, $(0.4875,0.5560,$ $-0.0786,0.00999)$.

In the same order, $F_J/F_R=(1.34, 1.06, 1.04,1.14)$, so Joule heating is more than sufficient to balance radiative cooling. The excess is balanced mainly by the negative contribution of $F_c$.

Joule heating is the primary source of the thermal energy that drives the radiative emission. As discussed in the next section, the primary driver of $F_R$ is the static electric field $E_y$, and most of $F_R$ is due to proton Pedersen current dissipation.  

\subsection*{4.5. Driver and Components of the Joule Heating Rate}

The Joule heating rate $Q_J= J_y (E_y+V_z B_x/c) = Q_P +Q_S +Q_T$. The first equality shows that $Q_J$ is driven by
the CM electric field, which is the sum of the static electric field $E_y$, and the convection electric field $V_z B_x/c$.
The relative importance of these two drivers is discussed in \S 4.5.1. The second equality shows that $Q_J$ has three components, due to different physical processes. The contributions of each one is discussed in \S 4.5.2, and the related  magnetization induced resistivity is discussed in \S 4.5.3.

\subsubsection*{4.5.1. Convection vs. Static Electric Field Driven Joule Heating}

Let $Q_{J,m}$ be the maximum value of $Q_J$ for a given sheet.
It is found that for all solutions the ratio $|V_z B_x/c E_y| \sim 10^{-4}-10^{-1}$
in the region where $Q_J \geq 10^{-2} Q_{J,m}$, and is $\sim 10^{-3}-10^{-1}$ in the region where $Q_J \geq 
10^{-1} Q_{J,m}$. For all
sheets, the smallest values of the ratio are in the regions where $Q_J$ is a maximum. Then it is $E_y$ that directly drives 
$Q_J$. Part of the reason for this is the relative smallness of $V_z$. As shown
in figure 8, $V_z \sim 3-10$ m-s$^{-1}$ in the regions that emit essentially all of $F_R$. This is 2-3
orders of magnitude smaller than characteristic sound and MHD wave speeds, and expected field-aligned 
flow speeds in the lower and middle chromosphere.

\subsubsection*{4.5.2. Components of $Q_J$}

Figures 9 and 10 show the components of $Q_J$ along with $Q_R$ for comparison. Define the Spitzer, Pedersen, and
thermoelectric heat fluxes $F_S,F_P$, and $F_T$ as the integrals of $Q_S, Q_P$, and $Q_T$ over $z$. Then $F_J=F_S+F_P+
F_T.$

In order of decreasing $n(0)$ the values of these fluxes for each of the four sheets, in units of $10^6$ 
ergs-cm$^{-2}$-s$^{-1}$, are $(F_R,F_J,F_P,F_S,F_T)=$ 
$(4.451,5.974,3.694,2.282,-0.0012)$, $(1.025,1.085,0.779,$ $0.316,-0.0105)$, 
$(1.55,1.604,1.209,0.486,-0.0913)$, $(0.488,0.556,0.502,$ $0.153,-0.099)$.

In the same order, the fraction of the resistive heating due to Pedersen current dissipation is $F_P/(F_P+F_S)=(0.6181,0.7113,0.7131,0.7662)$.
This shows that Pedersen current dissipation makes the dominant positive contribution to $F_J$ ($F_T$ makes a negative contribution), and hence is the main source of the thermal energy that drives radiative loss. The variation of $F_P/(F_P+F_S)$ with $n(0)$ suggests that Pedersen current
dissipation becomes increasingly dominant with increasing height in the chromosphere, consistent with the expected increase of the magnetization induced resistivity $\Gamma \eta$ with increasing height.

$F_T$ makes a small to moderate negative contribution to $F_J$ that in magnitude is $\sim 0.02 - 15 \%$ of the positive contribution of $F_P+F_S$ to $F_J$. This thermoelectric cooling effect is largest in the lowest density current sheet.
This sheet has the largest temperature gradient. 

\subsubsection*{4.5.3. Magnetization Induced Resistivity}

The total resistivity $(1+ \Gamma) \eta$, and the amplification factor $1+ \Gamma$ are shown in figure 11. 
In the thin regions that emit almost all of $F_R$, the resistivity decreases by 2-3 orders of magnitude
from the exterior to the interior of these regions. Since $\rho_H/\rho \sim 1$, and $\Gamma = (\rho_H/\rho)^2 M_e M_p$, the 
figure shows that this variation is entirely due to the decrease in magnetization. As the neutral sheet is approached
$1+\Gamma \rightarrow 1$ since the magnetization factor $M_e M_p$ tends to zero. Since, as indicated in \S 4.5.2, 
$Q_P (= \Gamma \eta J^2)$ comprises $62-77 \%$ of the positive contribution to $Q_J$, it follows that the magnetization induced resistivity $\Gamma \eta$ plays a major role in generating $Q_J$, which drives the radiative loss. If $\Gamma$ is set equal to zero, the current sheet solutions presented here do not exist.

\subsubsection*{4.6. Total Energy Conservation and Net Energy Fluxes}

Denote the net enthalpy, bulk kinetic energy, gravitational potential energy, ionization energy, and Poynting fluxes into the slab $0 \leq z \leq H = 50$ km by $F_H, F_{KE},F_g,F_{ion}$, and $S_z$. They are respectively the differences
between the values of $5 p V_z/2, \rho V_z^3/2, - \rho g R_\odot V_z, \chi_H n_e V_z$, and $- c E_y B_x/4 \pi$ at $z=0$ and $z=H$, where $R_\odot$ is the solar radius. For example $F_H = 5(p(0)V_z(0)-p(H)V_z(H))/2$. Here $F_g=0$ since 
$\rho V_z$ is constant. Define the total non-radiative flux $F_{NR}=F_H + F_{KE}+F_{ion}+S_z$. 

Combining the momentum and thermal energy equations with Poynting's theorem gives the total energy equation. It states
that the divergence of the total energy flux is zero. Integrating this equation over the slab gives $F_{NR}= F_R$. 

For the four solutions in order of increasing $n(0)$, and in units of $10^6$ ergs-cm$^{-2}$-s$^{-1}$, the net fluxes are as follows: $(F_H, F_{KE},F_{ion},S_z, F_R)=$
$(-0.1924,-0.7906 \times 10^{-3},-1.2571 \times 10^{-2},4.7572, 4.4507)$,
 $(-0.309 \times 10^{-2}, -2 \times 10^{-7}, -0.427 \times 10^{-3},1.043,1.0253)$, 
$(1.64 \times 10^{-3},-3.8 \times 10^{-6}, -0.3264 \times 10^{-2},1.5601,1.5499)$, and 
$(-0.01183,-3.2161 \times 10^{-3},0.4383 \times 10^{-2}, 0.4816,0.4875)$. 
Since $S_z \sim 20 - 1000$ times larger than any other non-radiative flux, it is
essentially the net upward Poynting flux into the slab that drives the heating and corresponding radiative loss.  

For the four solutions in order of increasing $n(0)$, $F_{NR}$ equals $F_R$ with a relative numerical error of $2.26, 1.38, 0.55$, and $1.38 \%$. This is a check on the accuracy of the numerical solution.

\section*{5. Discussion and Conclusions}

The current sheet solutions presented here represent the first, first principles theoretical proof of the existence of radiating current sheets under chromospheric conditions, based on a self-consistent model that includes reasonably realistic descriptions of anisotropic electrical conduction and thermoelectric effects, and NLTE ionization and radiative cooling. The existence of these solutions suggests the existence of sub-resolution, horizontal current sheets in the chromosphere that are sites of strong Joule heating driven radiative emission in that no more than a few tens of the sheets collectively emit the total radiation flux from the chromosphere. 

Current observations cannot rule out the existence of such strongly radiating current sheets. As indicated in \S 4.2, the model predicts that essentially all radiative emission from the sheets comes from regions with thicknesses $\lesssim 1 - 10$ km. A spatial resolution $\lesssim$ a few km would be necessary to image individual sheets seen edge-on, and a resolution $\sim 10-100$ times smaller would be necessary to accurately determine the properties of the sheets. The current best images of chromospheric plasma have a much lower resolution $\sim 100$ km (e.g. van Noort \& Rouppe van der Voort 2006). But for random viewing angles most such sheets would be seen as much larger sheet structures with dimensions unspecified in the 1.5 D model used here. 

Even if such strongly radiating sheets exist, they cannot be the only significant source of chromospheric heating, for the following reason.
Their small thickness combined with the small number of sheets needed to balance the chromospheric net radiative loss imply the filling factor of
such a collection of sheets would be so small that their collective emission would be more inhomogeneous than observations of chromospheric emission indicate. Given the complexity of the real chromosphere relative to the model used here, the current sheet solutions presented here are broadly interpreted as evidence that significant chromospheric heating occurs in current sheets with a range of emission rates and filling factors \footnote{Work in progress shows that current sheet solutions with heating and emission rates $\sim 4$ times smaller than those of the solutions presented here exist, consistent with the possibility that significantly more than a few tens of current sheets are collectively the sites of significant chromospheric heating. Current sheet solutions with still smaller heating and emission rates might exist.}. 

The existence of the current sheet solutions presented here proves that the MHD model combined with collision dominated transport and NLTE ionization and radiative cooling contains the essential physics for current sheet formation under chromospheric conditions.
The current sheet solutions presented here arise spontaneously as a result of the intrinsic transport physics of the model in that there is no a priori specification of solutions to the model to cause it to generate solutions of this or any other type. These current sheets are not of Sweet-Parker type since the only component of the flow velocity with a gradient is in the direction perpendicular to the sheets, and since the mass flux normal to the sheet is constant.  

The horizontal current sheets modeled here might form as a result of magnetic bipoles rising into the chromosphere as
their footpoints separate, creating a region of horizontal magnetic field that may form a current sheet as it 
is forced into an overlying region with a magnetic field having a horizontal component of opposite magnetic polarity.
Since this bipole emergence and rising process occurs continually over the entire surface of the Sun, it might be
an important quasi-steady state driver of chromospheric heating.

The Joule heating rate $Q_J$ that drives the radiative loss in the current sheets is mainly due to proton Pedersen current dissipation, but electron current dissipation makes a significant contribution due to it being the dominant heating mechanism near the neutral sheet where the plasma is weakly magnetized. It is the absence of a guide magnetic field (i.e. $B_z=0$) that causes the magnetization to go to zero at the neutral sheet, causing heating by electron current dissipation determined by the Spitzer resistivity $\eta$ to dominate the heating rate near the neutral sheet. Introducing a guide field strong enough to maintain a state of strong magnetization $(\mbox{i.e.}\; M_e M_p \gg 1)$ throughout the current sheet would cause $Q_J$ to be
essentially entirely due to proton Pedersen current dissipation. Introducing a nonzero guide field significantly complicates the model, mainly by increasing the dimensionality of the input parameter space from five to nine. This is discussed in \S 6.2.  

The radiative cooling rate $Q_R$, and the degree of ionization are exponentially sensitive functions of $T$, as shown by the ionization and radiative cooling rate equations (\ref{saha}) and (\ref{Qrad}). Since $T$ is largely determined by $Q_J 
(\sim (1+\Gamma) \eta J_\perp^2)$, it follows that $Q_R$ and the degree of ionization are exponentially sensitive
functions of the resistivity $(1+ \Gamma)\eta$, especially its magnetization induced component $\Gamma \eta$ that gives rise to most of the thermal energy that drives $Q_R$. This is an example of why it is necessary to accurately model and numerically resolve transport processes to accurately predict radiative loss in the atmosphere. This argument applies to all potential drivers of chromospheric and coronal heating.

Since most of the heating in the current sheet solutions is due to Pedersen current dissipation, these solutions provide yet another example of how Pedersen current dissipation, in this case driven by what is viewed as a quasi-static induction electric field, can be an important chromospheric heating mechanism. There is a long standing proposition that Pedersen current dissipation is the main chromospheric heating mechanism, causing and maintaining the chromospheric temperature inversion. This proposition is discussed in \S 6.3.

\section*{6. Further Discussion}

\subsection*{6.1. Determining the Input Parameters}

The input parameter space for the model presented here is five dimensional. There does not appear to be any way to know a priori
what sets of input parameter values lead to physically reasonable solutions. After choosing values for $p(0),
B_x(0)$, and $T(0)$ based on current knowledge of chromospheric conditions, the following three heuristic guidelines are used to constrain the search for $E_y$ and $V_z(0)$, resulting in the discovery of the current sheet solutions presented here, and a set of solutions that do not contain current sheets: 
\begin{enumerate}
\item If $B_x(0)>0 (<0)$, a necessary condition for the solution to contain a current sheet is that the constant field 
$E_y$ be $<0 (>0)$. The reason is that this is the condition for the vertical Poynting flux to point into the current sheet on
both sides. This ensures that electromagnetic energy flows into the sheet to provide the energy for the Joule heating that drives radiative emission and ionization. 

\item Although $E_y$ is constant in the model, it is assumed that it represents a quasi-static field that is essentially
constant over an observed characteristic period $t_c \sim 600$ s of Doppler intensity oscillations in the chromospheric
network (Lites, Rutten \& Kalkofen 1993: the full range of observed periods is $\sim 5-20$ min). The $x$ component of Faraday's law gives $t_c |\dot{B_x}|\sim c |E_y| t_c/L$, where $\dot{B_x} \equiv \partial B_x/\partial t$, and $L$ is the scale height of $E_y$. It is required that $L \gg$ the height range of the solutions to the model since $E_y$ is assumed constant over this range. It is also required that $t_c |\dot{B_x}| \ll B_x(0)$ in order that the model assumption $\dot{B_x}=0$ be justified. Choose $L$ to be twice the ideal gas pressure scale height for $T=7000$ K. Then $L=422.2$ km, which is $\gg$ the 50 km height range over which the solutions are generated. The magnetic field variation 
$t_c |\dot{B_x}|$ is estimated as follows. Observed temporal variations in the intensity of the emission from the lower chromospheric network are typically $\sim 20 \%$ (Banerjee, Doyle \& O'Shea 1999). It is assumed this emission is
mainly due to the Pedersen current dissipation rate $\eta \Gamma J_\perp^2$. This scales with magnetic field strength $B$ as $B^4$, where a factor $B^2$ comes from the magnetization induced resistivity factor $\Gamma$, and a factor $B^2$ comes from $J_\perp^2$. The fractional change in intensity during the time $t_c$ is then assumed to be $\sim (t_c/B^4) \partial B^4/\partial t= 4 t_c \dot{B}/B$. Setting this equal to $20 \%$ gives $t_c \dot{B}/B \sim 0.05$. Based on this
estimate choose $t_c |\dot{B_x}| \sim 0.05 B_x(0)$. Then the preceding Faraday law estimate for $|E_y|$ gives $|E_y| \sim
0.05 B_x(0)L/c t_c \sim 3.52 \times 10^{-3} B_x(0)$ V-m$^{-1}$. Then for $B_x(0)=25-100$ G, $|E_y| \sim 0.09-0.35$ V-m$^{-1}$, which is, remarkably and without our prior knowledge, close to the values of $0.12-0.25$ V-m$^{-1}$ found for the current sheet solutions. 

\item Given the preceding estimate of $E_y$, an estimate for $V_z(0)$ is obtained by assuming there is no significant Joule heating at the lower boundary point $z=0$, in which case the Ohm's law is approximately the ideal MHD Ohm's law, giving $V_z(0) \sim - c E_y/B_x(0) \sim - 0.05 L/t_c \sim 35.2$ m-s$^{-1}$, using the preceding estimate of $E_y$. This value is comparable to the values of $4-46$ m-s$^{-1}$ found for the current sheet solutions. 
\end{enumerate}

These guidelines approximately define the region of the input space where the search for solutions was performed, and where current sheet solutions were found to exist. No solutions were found with $V_z(0)=0$ or $E_y=0$.\footnote{After the four current sheet solutions were obtained, another set of four solutions were found by using the same sets of values of $T(0), p(0), V_z(0)$, and $|B_x(0)|$ as for the current sheet solutions, but reversing the sign of $B_x(0)$. Because of the way $E_y$ is computed in the computer code, reversing the sign of $B_x(0)$ changes the values of $E_y$ to $(0.1120,0.1051,-0.0727,
-0.1651)$ V-m$^{-1}$. These solutions extend over a 260 km height range, are smoothly varying, do not contain current sheets, and are relatively weakly heated and weakly radiating. These solutions may be used, for example, as background states to self consistently estimate the time dependent heating and radiative emission rates, and ionization driven by linear MHD waves.}

\subsection*{6.2. Including a Guide Magnetic Field}

The constant guide field $B_z$ is set equal to zero to simplify the model. If $B_z \neq 0$ 
the model becomes significantly more complicated in the following way. If $B_z =0$, the $x$ component of the
Ohm's law reduces to $E_x=0$, consistent with the fact that $\nabla \times {\bf E}=0$ implies that $E_x$ is constant. If 
$B_z \neq 0$, the $x$ component of the Ohm's law becomes an additional differential equation of the model. Then the model
is over determined unless it is assumed that $B_y \neq 0$. If $B_y \neq 0$, it follows from the $x$ and $y$ components of the momentum equation that $V_x$ and $V_y$ are not constant. These components of the momentum equation can be analytically integrated to obtain $V_x(z)$ and $V_y(z)$ in terms of $B_x,B_y,B_z$, and boundary conditions at $z=0$. The number of input parameters required to determine the solution increases from five to nine if $B_z \neq 0$, since $B_z,B_y(0), V_x(0)$, and $V_y(0)$ must now be specified. Although a nonzero $B_z$ does not directly affect the current density since $B_z$ is constant, the fact that $B_y \neq 0$ implies that $J_x \neq 0$, which directly affects the Joule heating rate. Since there is variation only with respect to $z$, $J_z$ remains zero.

$B_y$ cannot be a guide field. The reason is as follows. Assume $B_y$ is
constant. Allow the constant $B_z$ to have any value. The $x$
component of the Ohm's law, which is $E_x=0$ for the model used in the paper, is now $E_x +(V_y B_z - V_z B_y)/c 
= \eta J_y(M_e B_z/B - \Gamma B_x B_y/B^2)$, where $E_x$ and $V_y$ are constant. This $z$ dependent equation
is an additional equation that must be satisfied. This equation over-determines the solution to the model, so in general
there are no non-trivial solutions. Then a nonzero $B_y$ must depend on $z$, and so cannot be a guide field.

A guide field prevents the magnetization, and hence the magnetization induced resistivity $\Gamma \eta$, from decreasing to zero
at the neutral sheet. An important question is whether a guide field has a significant effect on 
the radiative emission flux $F_R$. Since $F_R$ is driven mainly by resistive heating, which is partly determined by the
resistivity $(1+ \Gamma) \eta$, and since the presence of a guide field increases $\Gamma$, it is expected that the guide
field has a direct effect on $F_R$. Although the model outlined in the first paragraph of this section must be solved to obtain 
a reliable estimate of the effect of a guide field, a simple though crude estimate of the effect of a small guide field on $F_R$
is obtained as follows. Modify $\Gamma (\propto B^2 = B_x^2)$ by replacing $B^2_x$ by $B^2_x + (0.1 B_x(0))^2$, where $B_x(0)$ is 
the given boundary value of $B_x$. This ad hoc modification mimics the effect on the resistivity of a guide magnetic field that 
is one order of magnitude smaller than $B_x(0)$. Then as the neutral sheet is approached, 
$\Gamma \rightarrow$ a non-zero value instead of zero. Running the code with this modification leads to a set of 
current sheets as before, and to corresponding values of $F_R$ that differ from those for the solutions discussed in the paper by  
$(-0.04, -0.11, 0.09, -0.007) \%$, in order of increasing number density. These differences are insignificant, and are
consistent with the values of $\Gamma$ at the neutral sheets, which are $(0.1564, 0.1123, 0.0975, 0.0245)$. Since $\Gamma \ll 1$,
the resistivity in the neighborhood of the neutral sheets is $\sim \eta$, as in the absence of a guide field. Since $\eta$ is mainly a function of $T$,
and since $T$ in the neighborhood of the neutral sheets for the modified solutions differs by $\lesssim 4.15 \times 10^{-2} \; \%$ from its values for the
solutions discussed in the paper, it follows that the value of $\eta$ in this neighborhood does not change. 
These results suggest that the addition of a guide field that is a factor of 10 or more smaller than the asymptotic value of the field component that generates the current density does not have a significant effect on the emission rate.

\subsection*{6.3. Pedersen Current Dissipation and Chromospheric Heating}

Pedersen currents are mainly ion currents that flow along and are driven by ${\bf E}_{CM \perp}\equiv$ ${\bf E}_\perp + ({\bf V} \times 
{\bf B})/c$, where ${\bf E}_\perp$ is the component of ${\bf E} \perp {\bf B}$ (e.g. Mitchner \& Kruger 1973).

Osterbrock (1961), following similar work by Piddington (1956), presents an analysis of damping rates of linear MHD waves in the chromosphere due to charged particle-neutral collisions\footnote{In the language of this paper, this is resistive damping of currents driven by the electric field of the waves, and governed by the resistivity $(1+\Gamma)\eta$. This includes Pedersen current dissipation.}, and a scalar viscosity, and concludes that damping of these waves is not important for chromospheric heating. Due to observational limitations at that time, Osterbrock (1961) uses magnetic field strengths orders of magnitude smaller than those now known to exist, causing the heating rate due to charged-particle neutral collisions to be under estimated by orders of magnitude. Advances in observations since 1961, combined with the largely complete development of the theory of collision dominated transport in variably ionized plasmas (Braginskii 1965) allowed for the recognition through theoretical modeling that Pedersen current dissipation may be an important chromospheric heating mechanism. 

There is a longstanding proposition that Pedersen current dissipation provides essentially all the thermal energy that drives chromospheric radiative emission, and that it is the increase in the magnetization of the plasma with height from the photosphere that triggers the onset of heating by Pedersen current dissipation sufficient to cause and maintain the chromospheric temperature inversion (Goodman 2000, 2004a). 

Heating by Pedersen current dissipation is not effective in the weakly ionized, weakly magnetized photosphere because the magnetization induced resistivity $\Gamma \eta (= M_e M_i (\rho_H/\rho)^2 \eta) < \eta$ due to weak magnetization $(M_e M_i < 1)$, and it is not effective in the strongly ionized, strongly magnetized corona because $\Gamma \ll 1$ due to strong ionization $((\rho_H/\rho)^2 \ll 1)$. This heating mechanism can only be effective in the weakly ionized, strongly magnetized chromosphere, between the photosphere and corona, since it is only there where $\Gamma$ can be orders of magnitude greater than unity, causing the plasma to be highly resistive with respect to the flow of Pedersen currents, allowing for, but not guaranteeing significant resistive heating. 

Although $\Gamma \eta$ can take on values orders of magnitude larger than $\eta$ under chromospheric conditions (e.g. figure 11, and references cited in this section), a large resistivity does not by itself imply a significant heating rate. An outstanding question is what are the primary drivers of Pedersen current dissipation in the chromosphere, and are they sufficiently strong to cause significant chromospheric heating? Any process that generates an ${\bf E}_{CM \perp}$ can in principle drive a Pedersen current, and hence some level of heating by Pedersen current dissipation.
The initial modeling of driving by convection electric fields, generated by the steady bulk flow of plasma $\perp {\bf B}$ is presented in Goodman (1997a,b), and extended in Goodman (2004b). The initial modeling of driving by linear Alfv\'{e}n waves is presented in DePontieu \& Haerendel (1998) and DePontieu, Martens \& Hudson (2001). Their basic model is extended in Goodman (2011) to include the full effects of the inhomogeneous background atmosphere on the waves from the photosphere to the lower corona. The initial modeling of driving by slow magnetoacoustic
waves, and of Pedersen current dissipation in the shock layers of fast magnetoacoustic shock waves are presented in Goodman (2000, 2001), and Goodman \& Kazeminezhad (2010), respectively. 

Much additional work has been done in this area. Overall, the models indicate that, under conditions consistent with current understanding of the chromosphere, Pedersen current dissipation driven by several different drivers can provide all the thermal energy needed to balance the total net radiative loss from the chromosphere. What is needed most to determine the effectiveness of this heating mechanism are calculations of ${\bf J}$ based on observations of ${\bf B}$ in the chromosphere. Calculations of this type have been done for sunspot chromospheres (Socas-Navarro 2005a,b; also see comments on those calculations in \S 6.1.2 of Goodman (2011)). Such calculations are especially challenging since ${\bf J}(\propto \nabla \times {\bf B})$ involves differences of derivatives of components of ${\bf B}$, so any error in the measurement of ${\bf B}$ is multiply compounded in the calculation of ${\bf J}$.    

\begin{acknowledgements}
The authors thank the referee for many comments that significantly improved the manuscript.
MLG gratefully acknowledges support from grants ATM-0650443 and ATM-0848040 from the Solar Physics Program of the National Science Foundation to the West Virginia High Technology Consortium Foundation. Part of this work was done while he was a visiting scientist at the High Altitude Observatory (HAO) of the National Center for Atmospheric Research (NCAR) in Summer 2011. He thanks the HAO staff, especially Philip Judge, BC Low, and Roberto Casini for their generous hospitality and scientific discussions. 
This research made use of NASA's Astrophysics Data System (ADS).  PGJ is grateful
to NASA for support of this work through award 09-LWSTRT09-0058 from the Living with a Star Targeted Research and
Technology program.

\end{acknowledgements}

\newpage
\appendix

\section{Effectively thin radiation loss function}

Radiation loss functions $\Phi(T)$ in units of 
erg~cm$^3$~s$^{-1}$, 
have been computed for decades (e.g., Cox \& Tucker 1969).  When multiplied by
electron and hydrogen number densities, they 
give the power per unit volume, $P$,
radiated by an optically thin plasma, given a set of elemental
abundances, when the plasma is at a sufficiently low density that only two-body collisions are important,
but high enough so that sufficient time has elapsed to reach ionization equilibrium:
\begin{equation}
  \label{eq:radp}
  P = \Phi(T) n_e (n_H+n_p)  \ \ {\rm erg~cm^{-3}~s^{-1}}
\end{equation}
Such functions are widely applied in coronal physics. 
But we need a radiation loss function for the chromosphere, which is in many 
transitions far from optically thin.  Yet it is possible to approximate the 
radiation losses exactly in the form of eq.~(\protect\ref{eq:radp}), with some care.  Radiative transfer
affects the radiation loss rates from the chromosphere in two important ways: first, some of the lines become effectively
thick, - in such regions these lines must not contribute significantly to the radiation losses; second, 
the ionization of hydrogen, which in a hydrogen dominated plasma determines $n_e$, $n_H$ and $n_p$, 
depends on radiation transport in the hydrogen lines and continua.   Both of these can be dealt with
following the work of Anderson \& Athay (1989) and Athay (1986) respectively.

Anderson \& Athay (1989) compute radiation losses in detail, using 
transfer calculations in the abundant emitting species. They 
show, in their fig.~9, that the function $f(T) \equiv P(T)/(n_p+n_H)$ 
varies only by a factor of two in quite different models.  Furthermore, it 
follows the form of $n_e/n_H$, also plotted by them, in regions where the plasma is partially ionized
($n_e \ll n_H$), i.e. $n_p+n_H \approx n_H$.  This figure then demonstrates
that the radiation losses in their chromospheric models are reproduced 
by better than a factor of two, by a formalism of the form of eq.~(\ref{eq:radp}).  This is a case where
Nature has been kind: while the chromosphere is a difficult, scattering-dominated regime of radiative 
transfer, the combination of efficient radiating ions (Fe II, Ca II, Mg II) is such that the 
effectively thin approximation works. All these species belong to dominant stages of ionization in the chromosphere 
and the scattering does radically affect each individual ion where it is expected to dominate.  The 
bottom line is that the radiation losses can be computed as effectively thin, i.e. the {\em optically} thin
approximation can be used. 

But there are potential problems with this approach. 
Athay (1986) examined the effects of the inclusion of H L$\alpha$ to genuinely
optically thin conditions, versus those that must exist in the upper regions of the chromosphere. 
The difference is essentially that, under optically thin conditions, photo-ionization of hydrogen 
by the Balmer continuum photospheric radiation is negligible, whereas under chromospheric conditions, it is significant.
The net result is that, compared with optically thin loss functions, the ``L$\alpha$ peak'' near $2\times10^4$ K
is much reduced, since before L$\alpha$ can be emitted at such temperatures, in optically thicker
plasmas such as the chromosphere, the otherwise neutral atoms are ionized by the Balmer radiation field from
the $n=2$ levels whose populations are enhanced by the scattering of photons that are
optically thick in the line core.     The proper solution of this problem requires a nLTE 
transfer calculation, in which transport of radiation  (neglected here) 
is calculated to yield the global coupling of the source function $S$, mean intensity $J$, and then
evaluation of $J-S$.  Fortunately, in our calculations 
the temperatures are low enough that the L$\alpha$ radiation never contributes
highly to our solutions.  

Another potential problem is that we neglect absorption of radiation in the energy equation.
But the sheets we have computed are optically thin to most radiation, 
so that if the internal plasma temperature of the sheet is above the equivalent radiation temperature
of radiation from the surrounding plasma (set for example, by light emitted at a far deeper layer of
the atmosphere such as the photosphere), then this is not a serious problem.  Given that
the temperatures we derive are higher than the photosphere of the Sun, we believe that 
our calculations are accurate enough. So finally, we computed the function $\Phi(T)$ using the following data sources. 
For ionization balance, we assume iron, calcium and magnesium, the dominant 
radiators, are singly ionized through the chromosphere- this is 
a valid approximation (Vernazza, Avrett \& Loeser 1981). Otherwise we
use the 
ionization and recombination 
coefficients of Arnaud+Rothenflug (1985) and compute 
two-body recombination rates 
using the OPACITY project cross sections, which includes dielectronic 
recombination through the inclusion of the resonances in photoionization 
cross sections.
(Seaton 1987).     A comparison of typical recombination 
rates computed from the OPACITY project and more detailed work 
(Nahar \& Pradhan 1992) is given in Judge (2007, fig 13): the agreement in recombination rate
coefficients is
within a factor of two.  Such uncertainties propagate to uncertainties in the radiation losses from
each element which are $\le$ a factor of two, and at any temperature, more than 
one ion contributes to the losses.  

For each radiating ion, the optically thin radiation losses depend primarily
on the inelastic collision rate with electrons. to compute these rates, we have 
adopted the CHIANTI database of 2006 (Landi et al. 2006), 
supplemented with Ca~II data which are 
important in the  chromosphere.    Radiation losses were computed for 
abundant elements, and the sum of all these losses was fitted to the form of 
 eq.~(\ref{Qrad1}).  This functional form fails above $1.5 \times 10^4$K.

\appendix

\input figs

\end{document}

%% file: figs.tex
\newpage

\newcommand{\pgjfig}[4]{
\epsscale{1}
\begin{figure}[!ht] 
\plottwo{#1}{#2} \caption {#3 (left), #4 (right).}
\end{figure}
}

\newcommand{\opgjfig}[2]{
\epsscale{.5}
\begin{figure}[!ht] 
\plotone{#1}\caption {#2}.
\end{figure}
}

\pgjfig{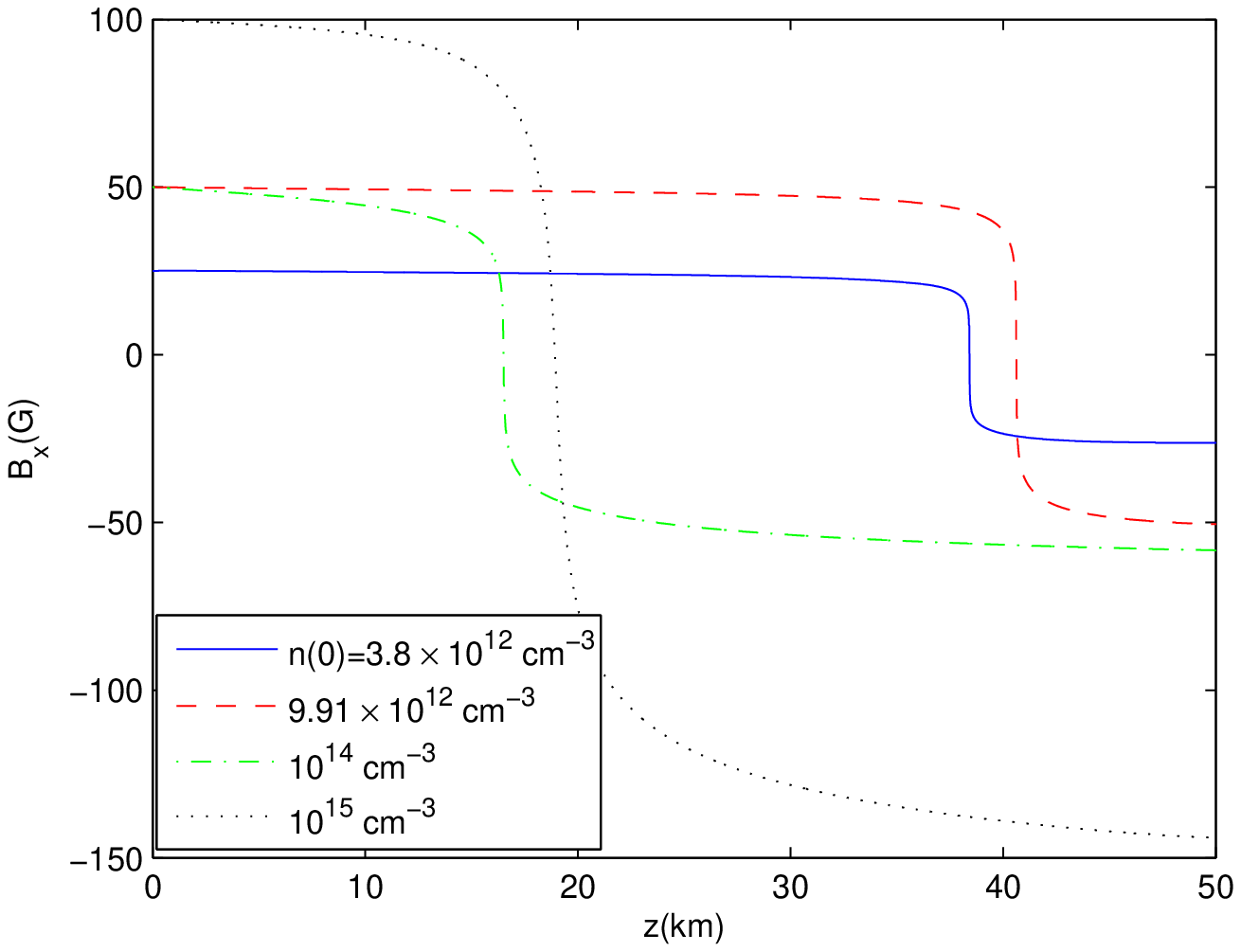}{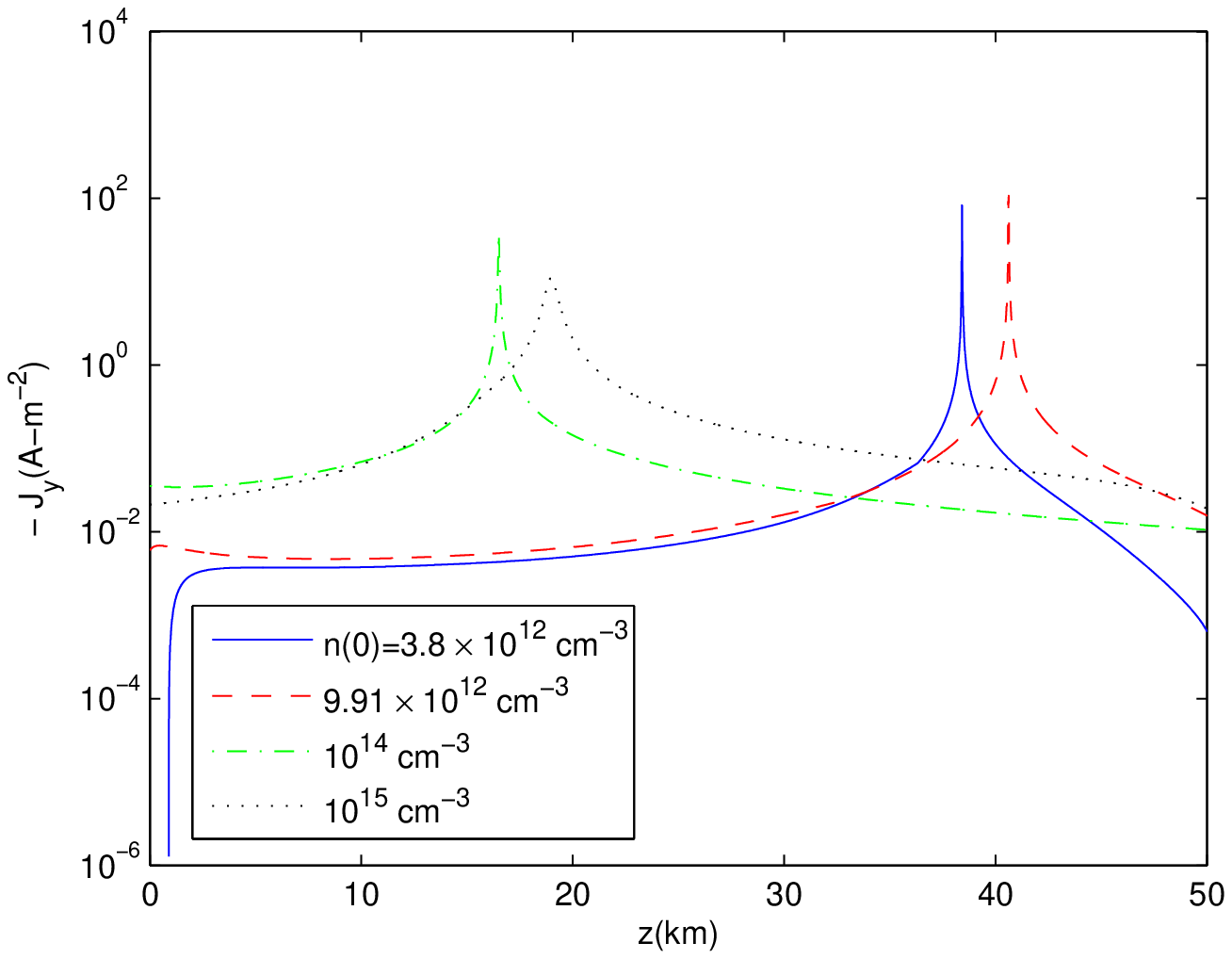} {Magnetic field vs. height}{current density vs. height}

\pgjfig{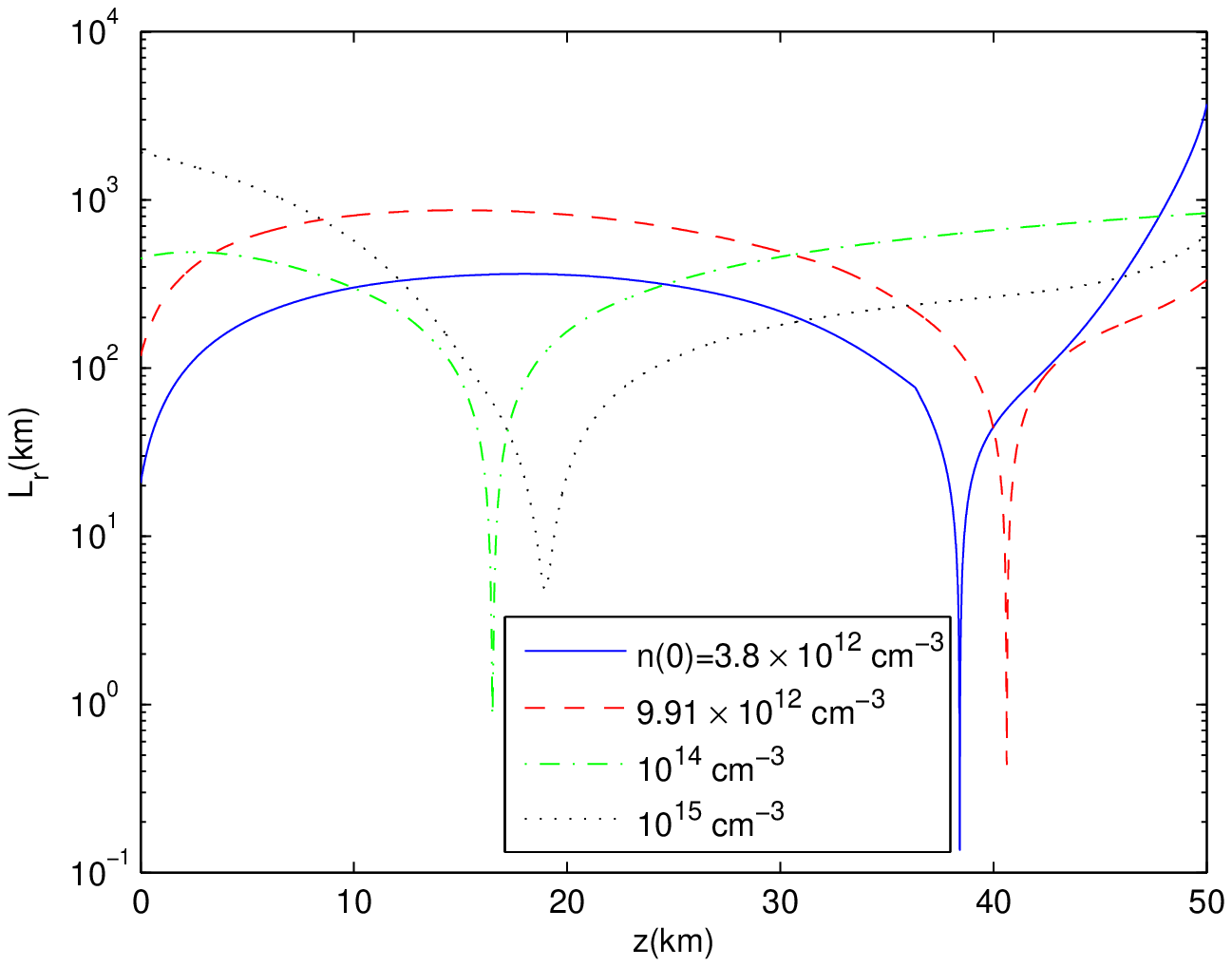} {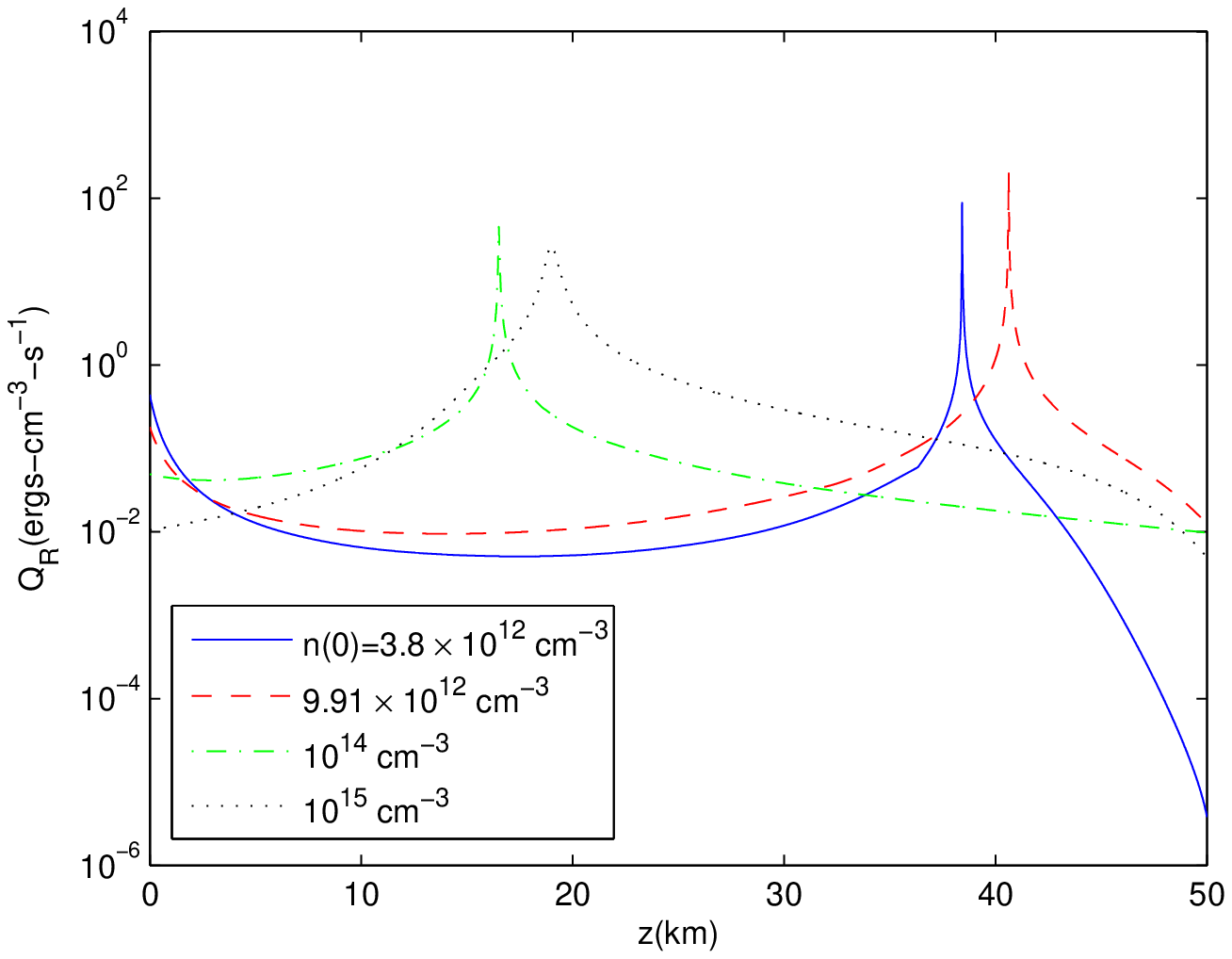} 
{Resistive length scale (equation \ref{reslengthscale}) vs. height}
{radiative cooling rate (equation \ref{Qrad}) vs. height}

\pgjfig{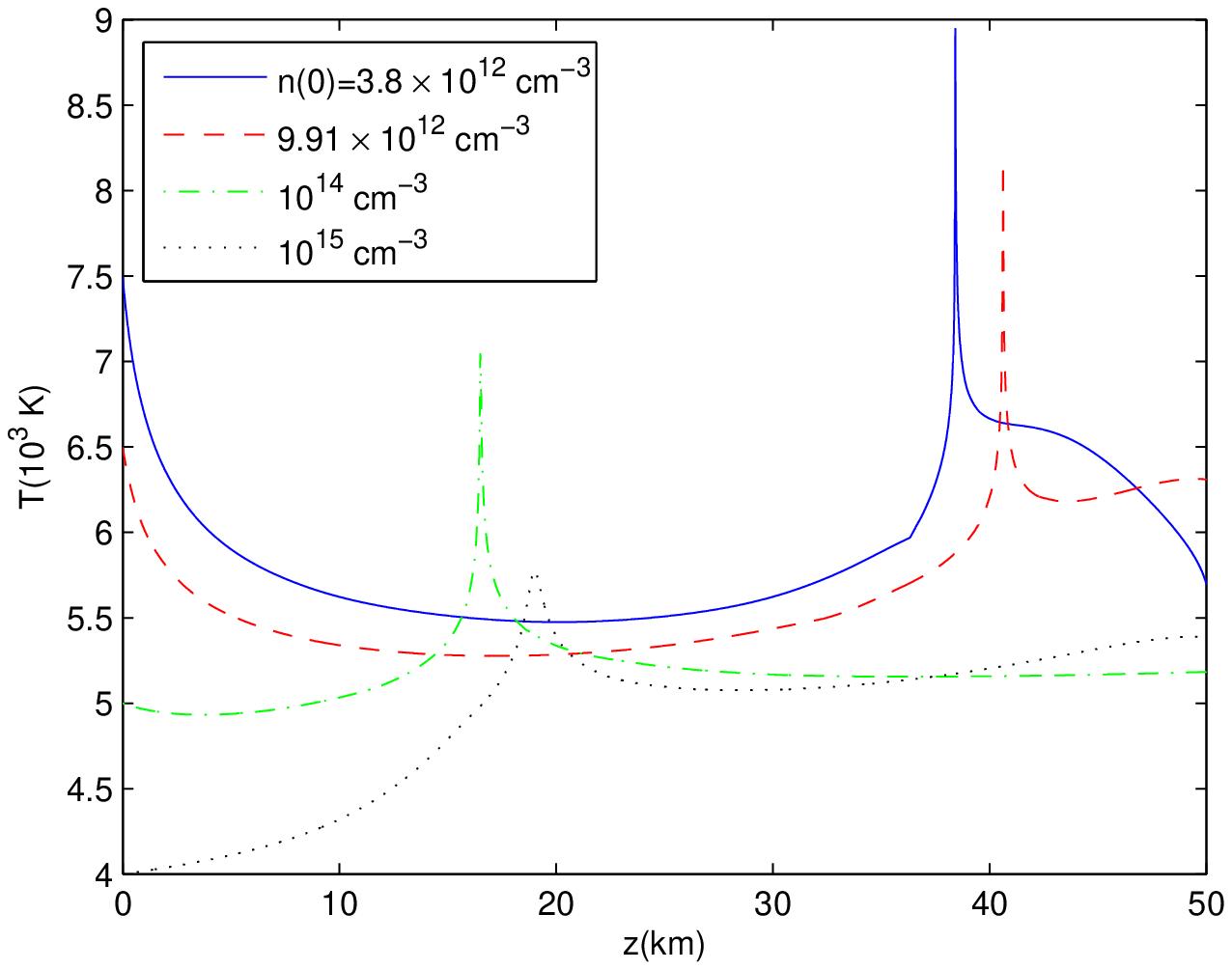} {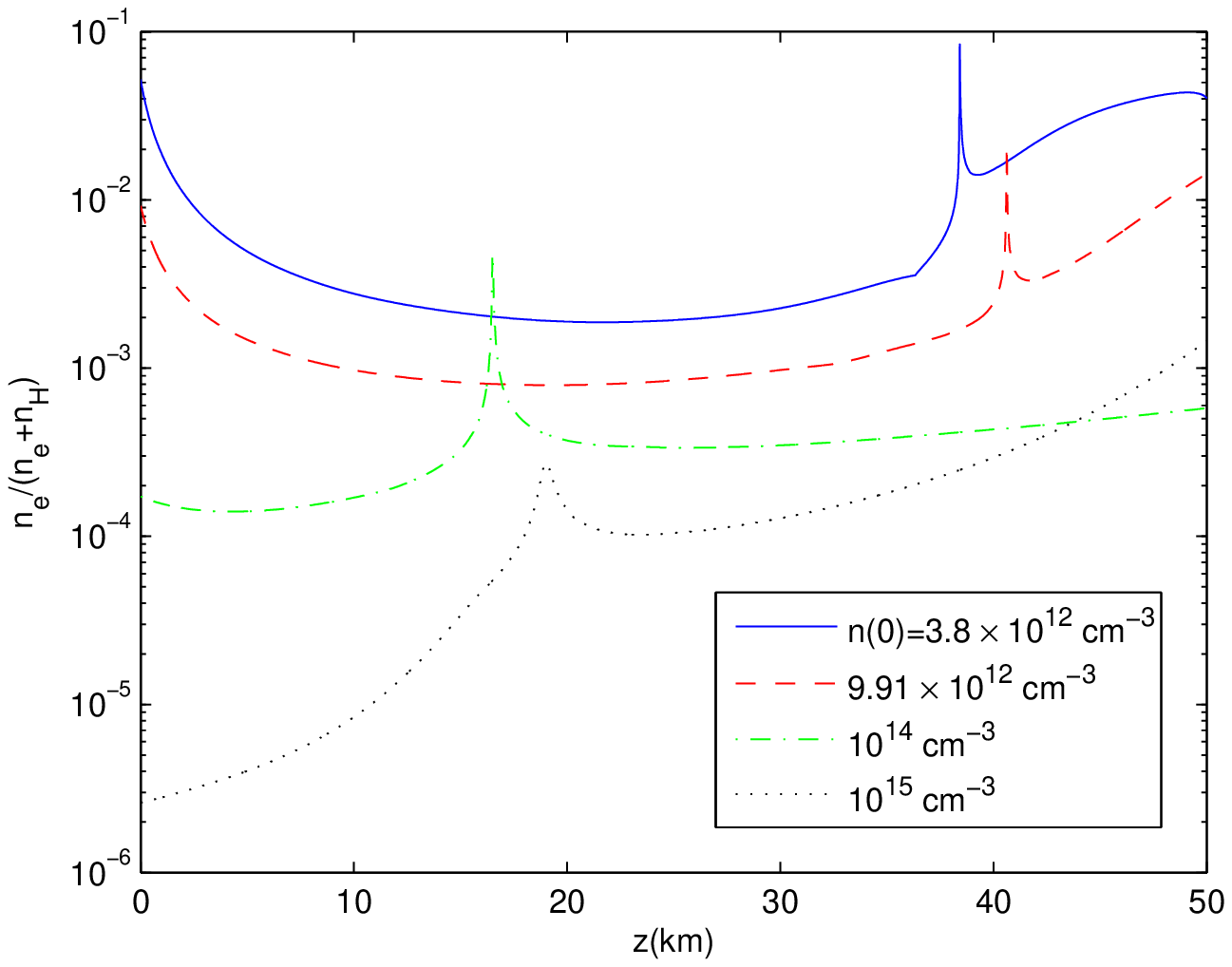}
{Temperature vs. height}
{ionization fraction vs. height}

\pgjfig{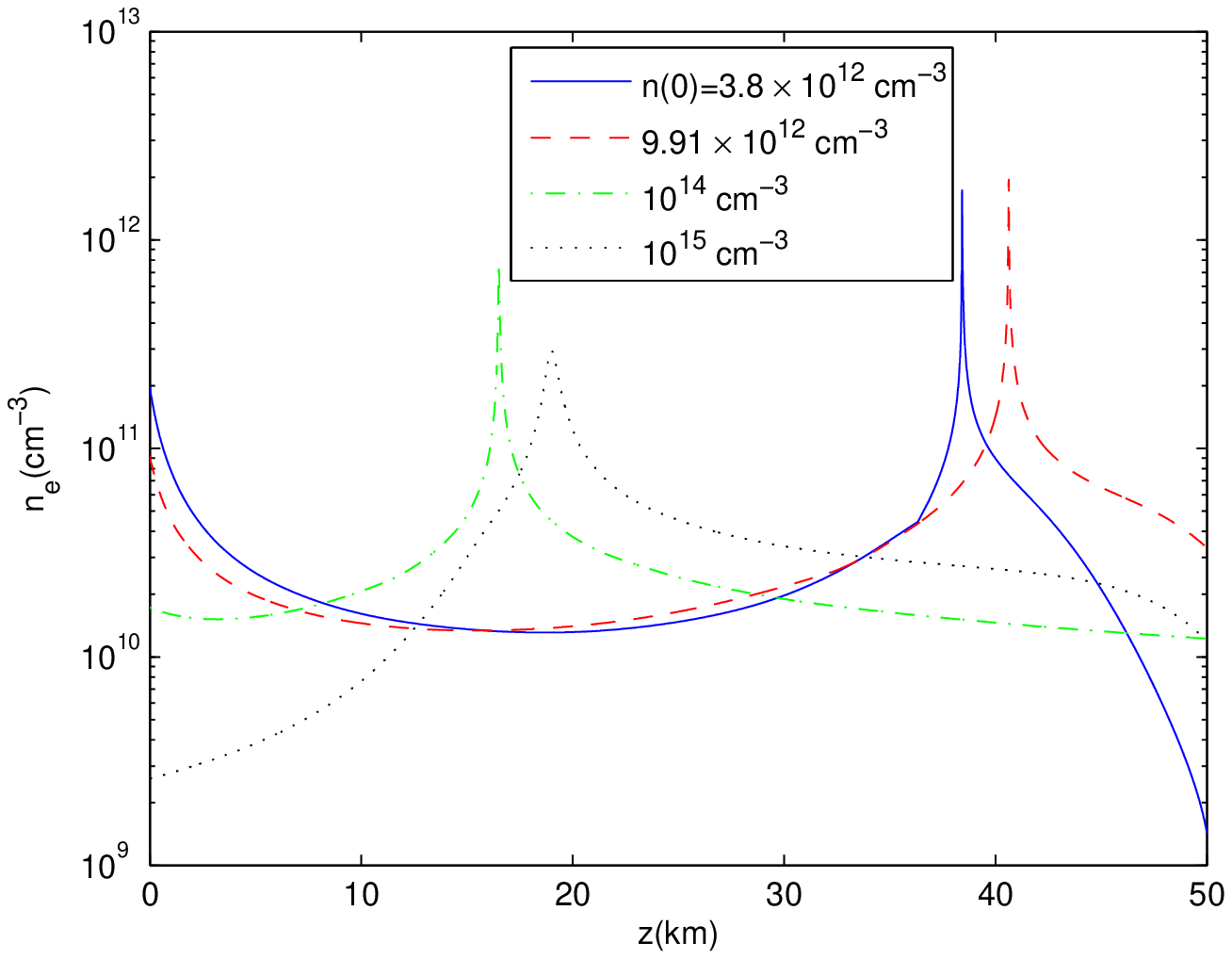}{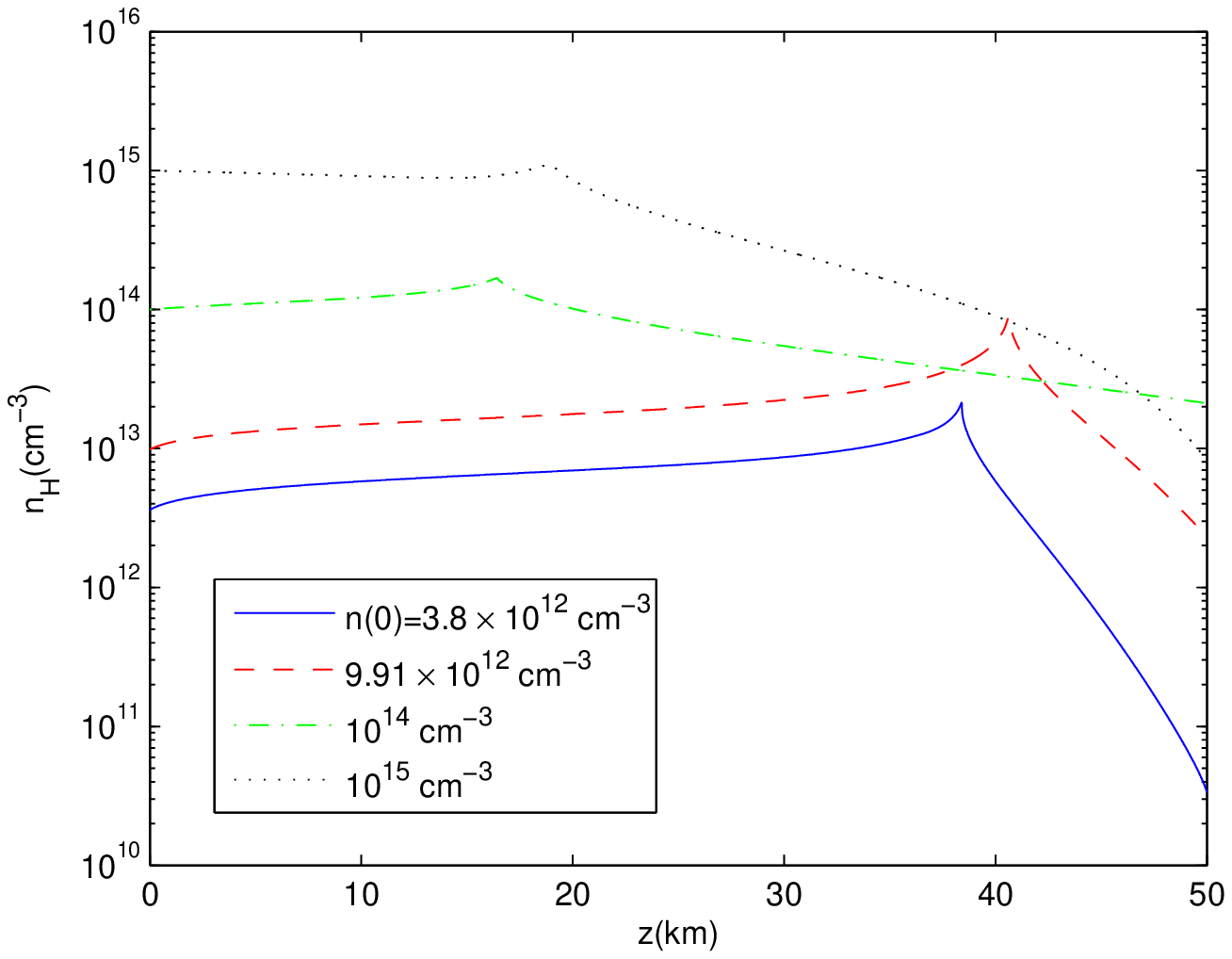}
{Electron number density vs height}
{HI number density vs. height}

\opgjfig{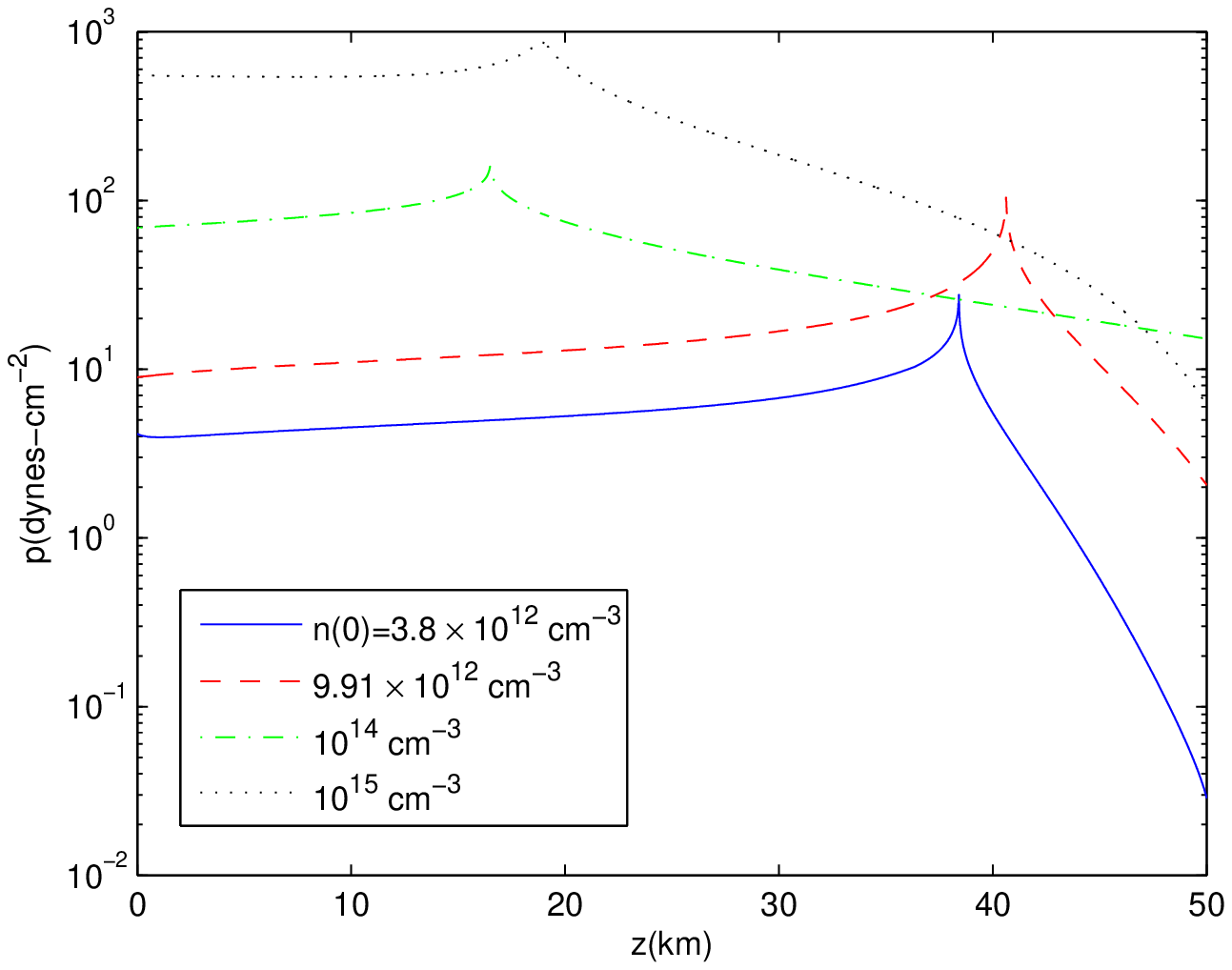}{Pressure vs. height.}

\pgjfig{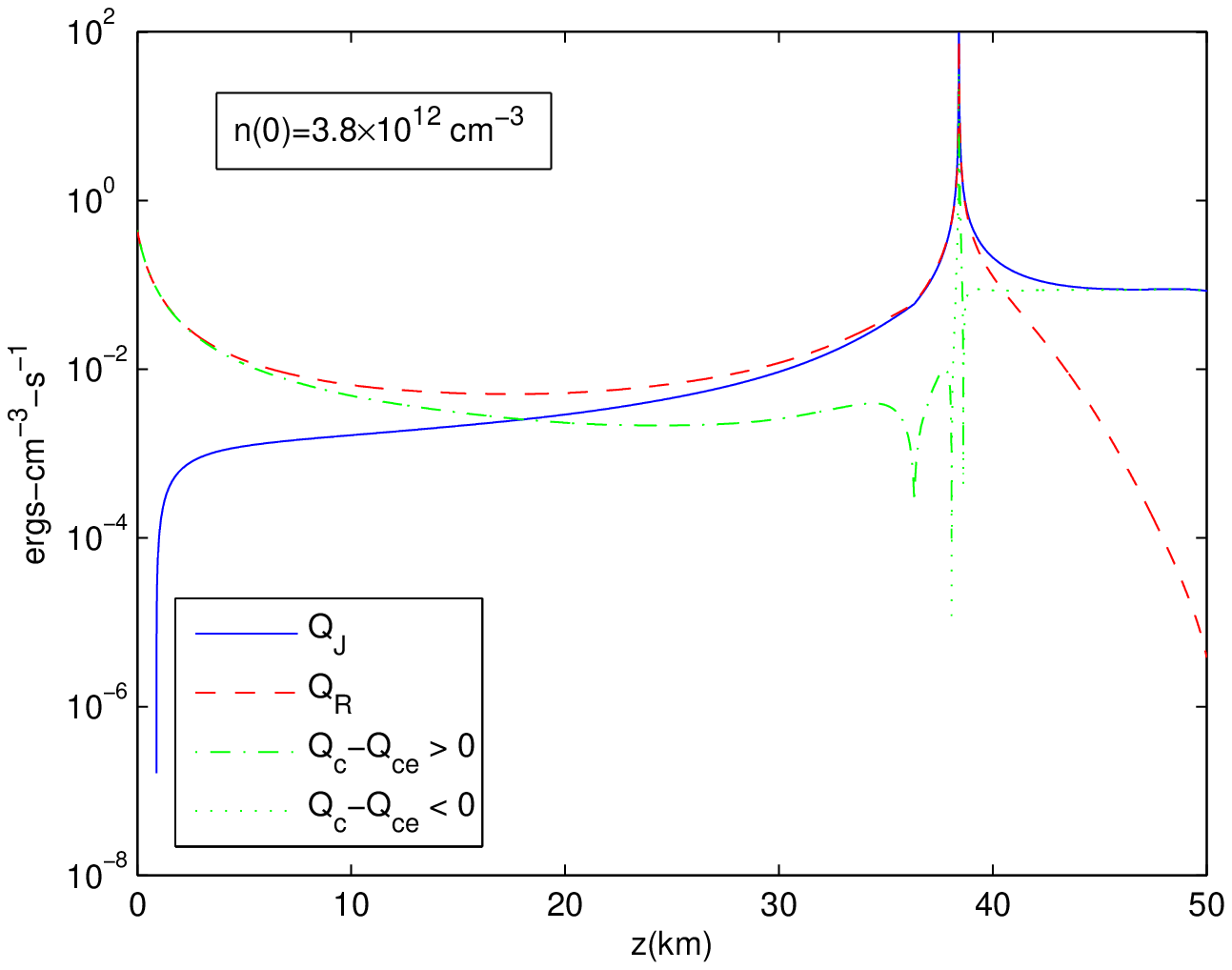}{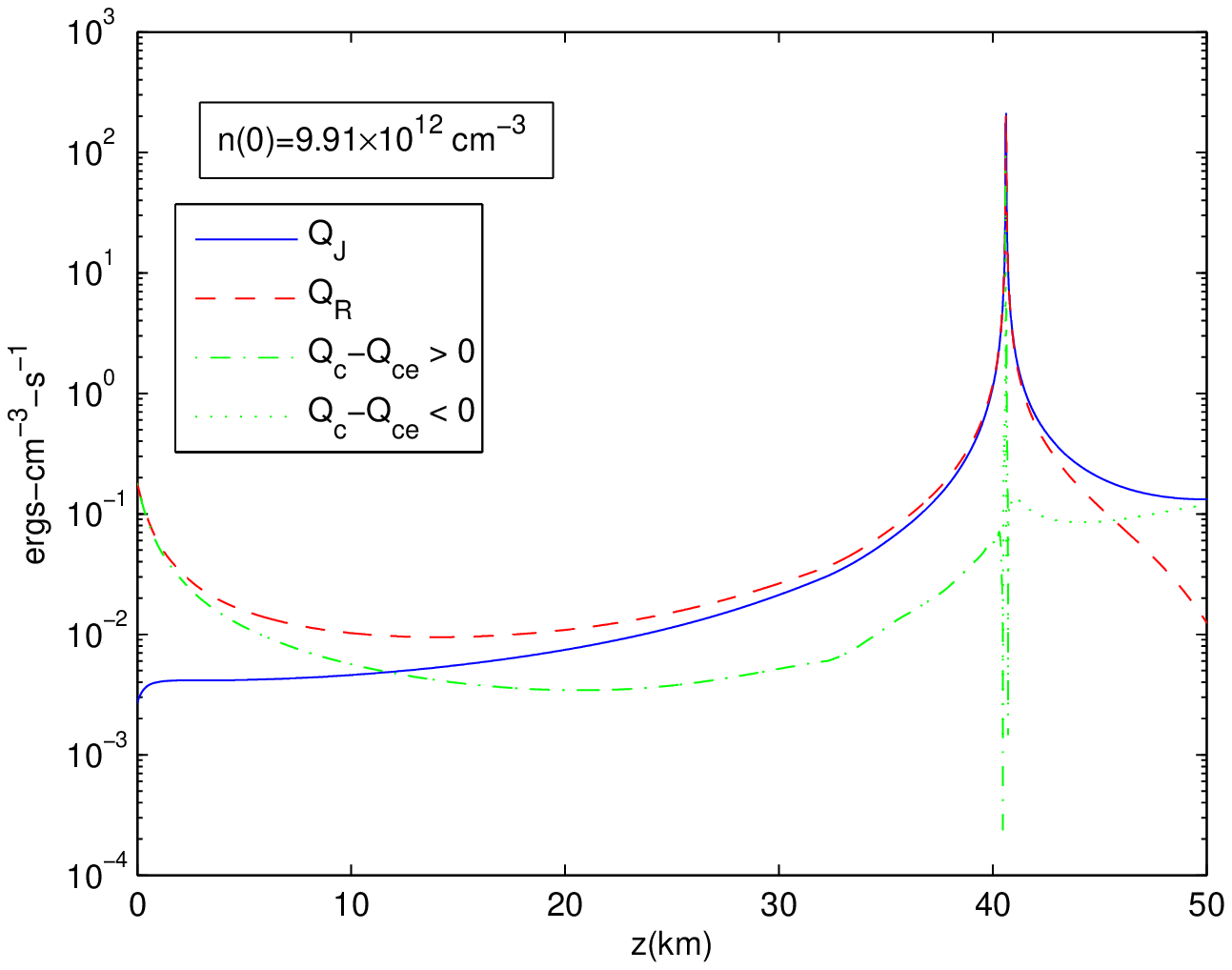} 
{Terms in the thermal energy equation (\ref{energy}) vs. height for the solution with $n(0)=3.8 \times 10^{12}$ cm$^{-3}$}{and for the solution with 
$n(0)=9.91 \times 10^{12}$ cm$^{-3}$}

\pgjfig{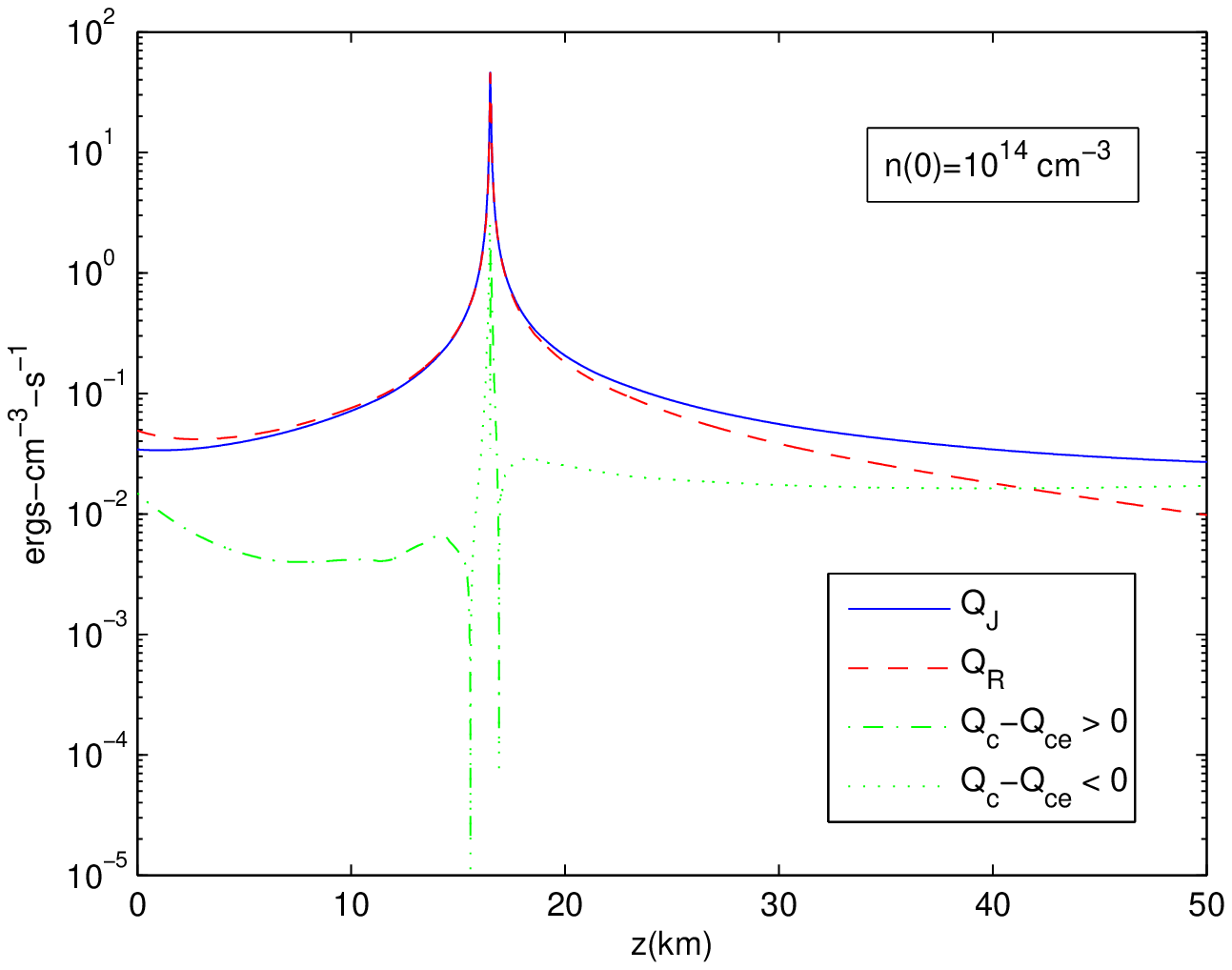} {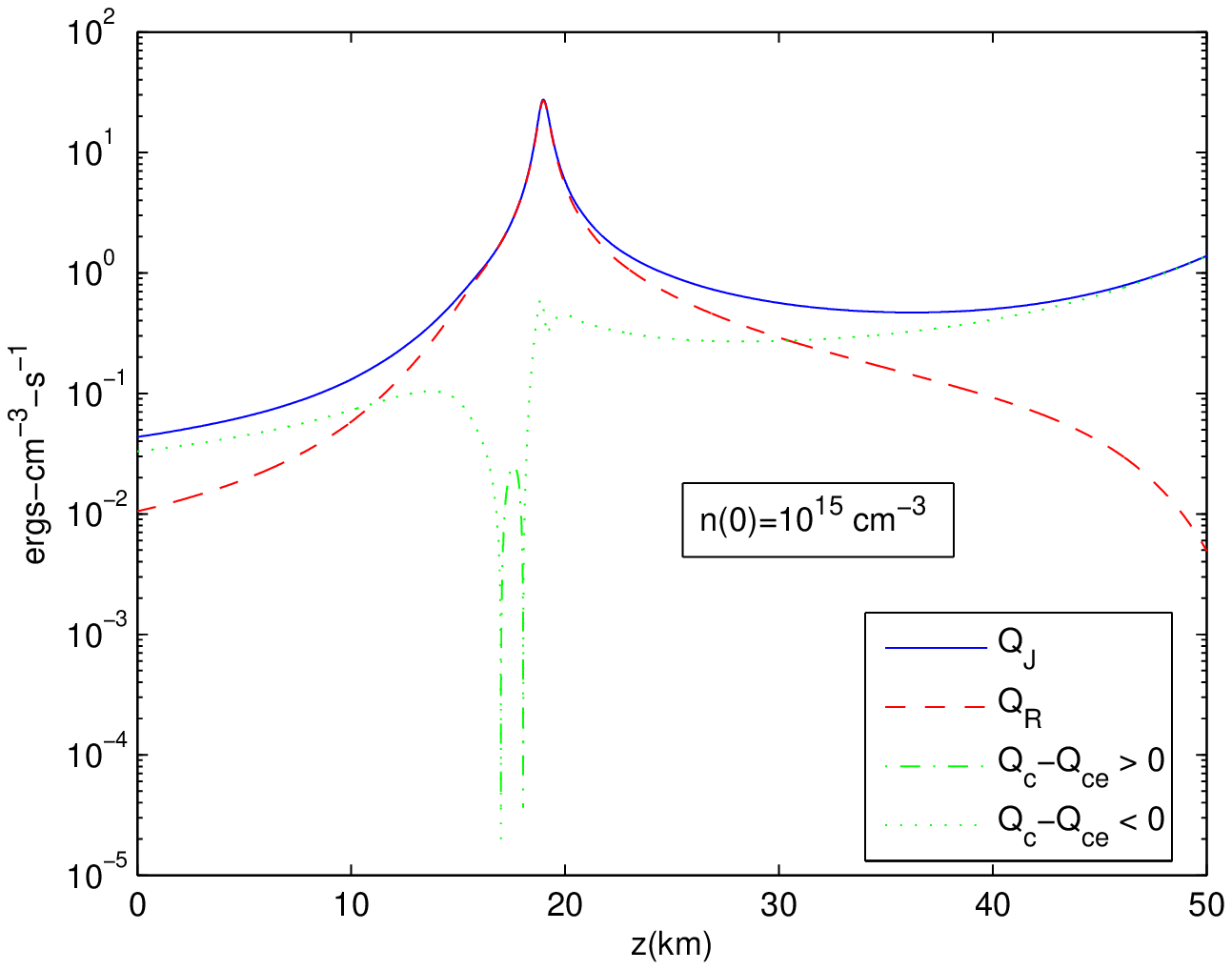} 
{Terms in the thermal energy equation (\ref{energy}) vs. height for the solution with $n(0)=10^{14}$ cm$^{-3}$}{and for the solution with $n(0)= 10^{15}$ cm$^{-3}$}

\opgjfig{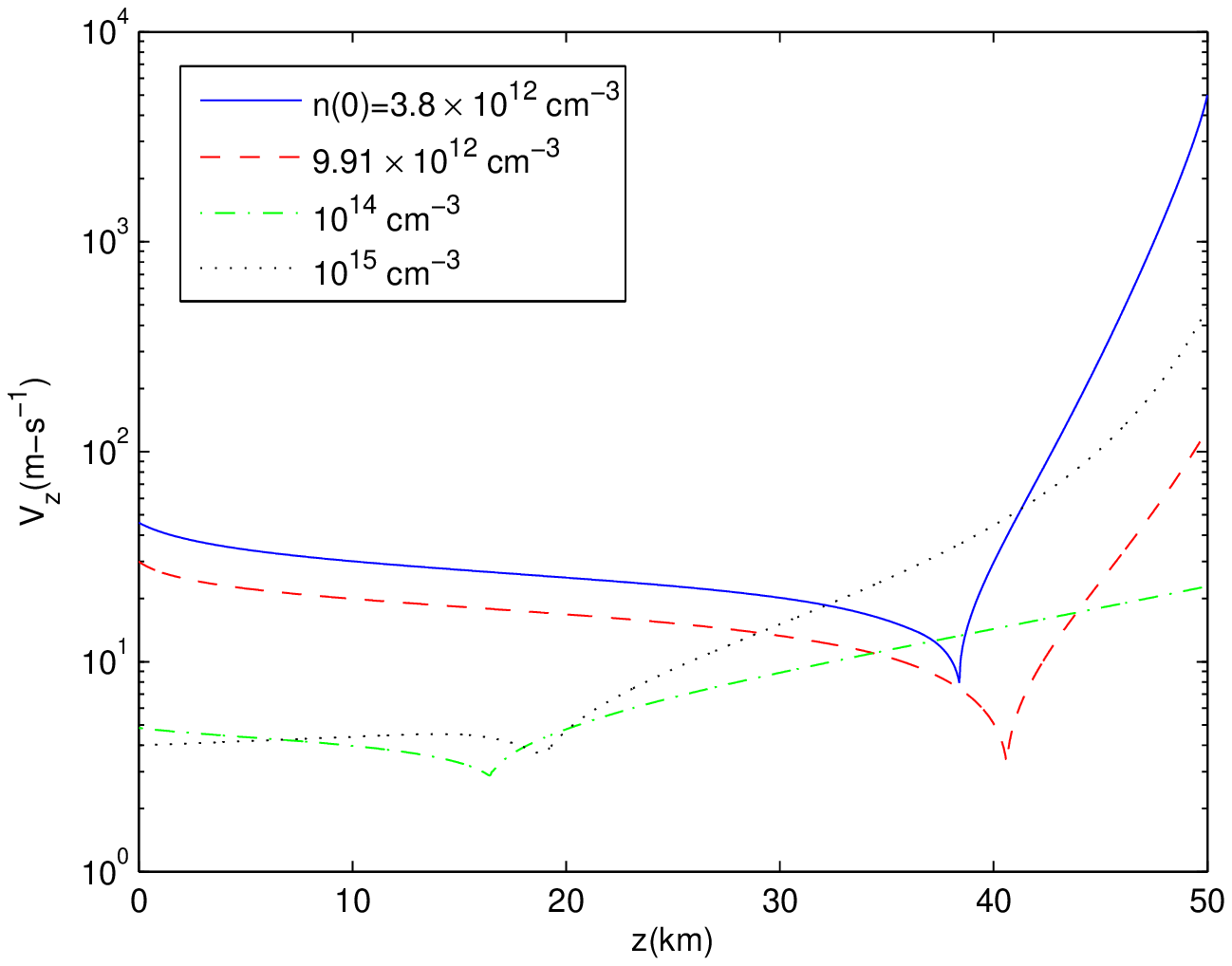}{Center of mass flow velocity vs. height.}

\pgjfig{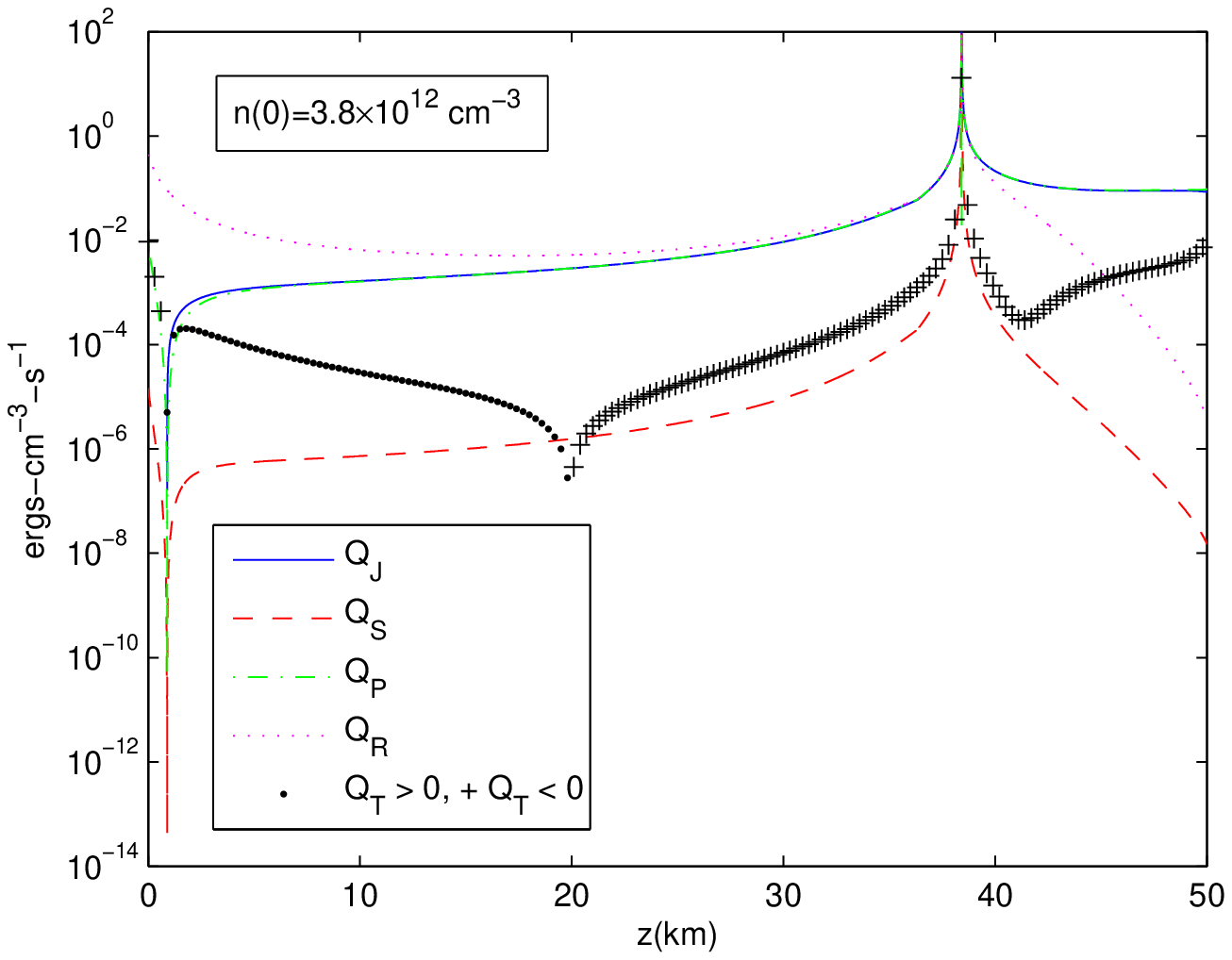}{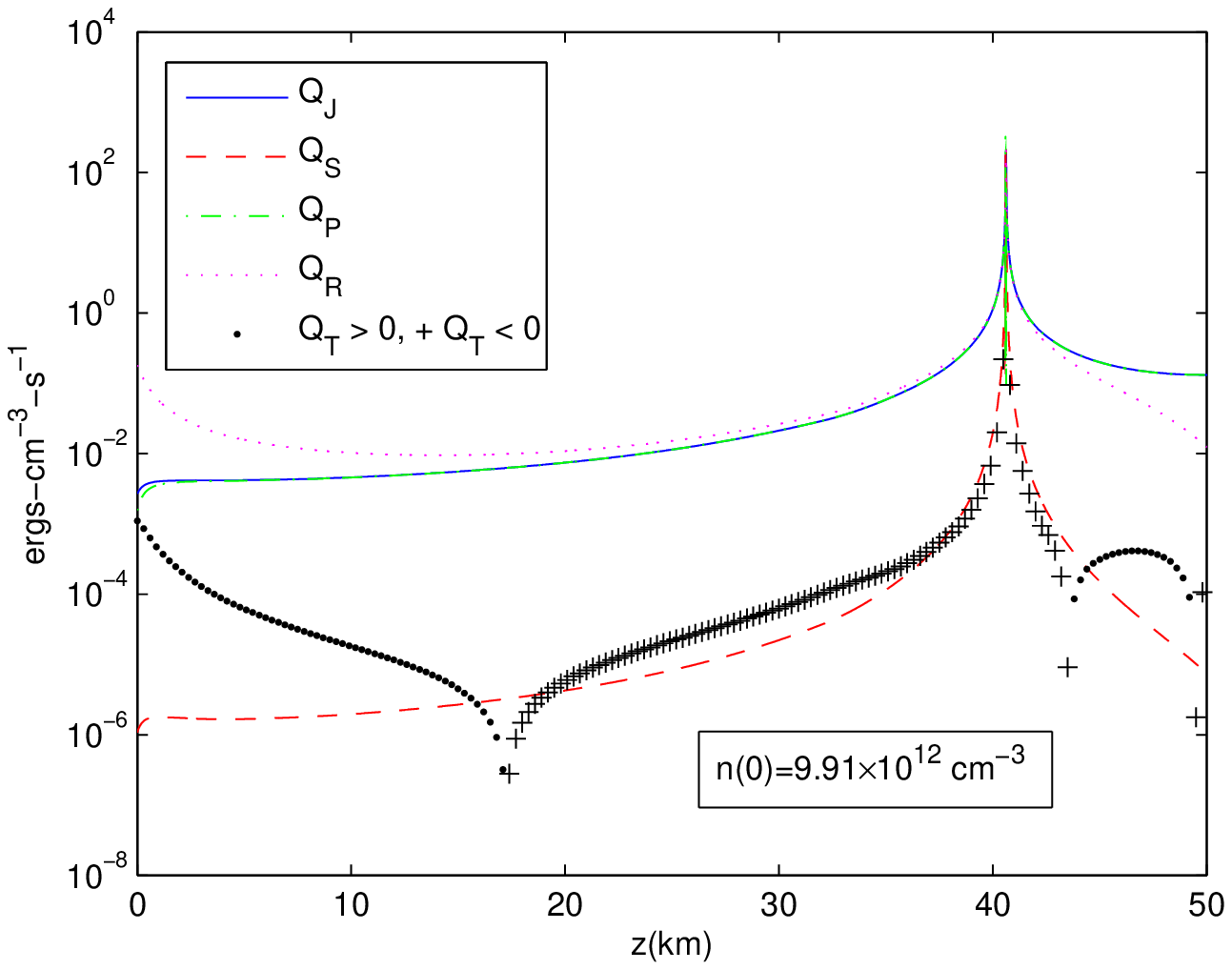} 
{Joule heating rate and its components, together with the radiative cooling rate vs. height for the
solution with $n(0)=3.8 \times 10^{12}$ cm$^{-3}$}{and for the solution with $n(0)=9.91 \times 10^{12}$ cm$^{-3}$}

\pgjfig{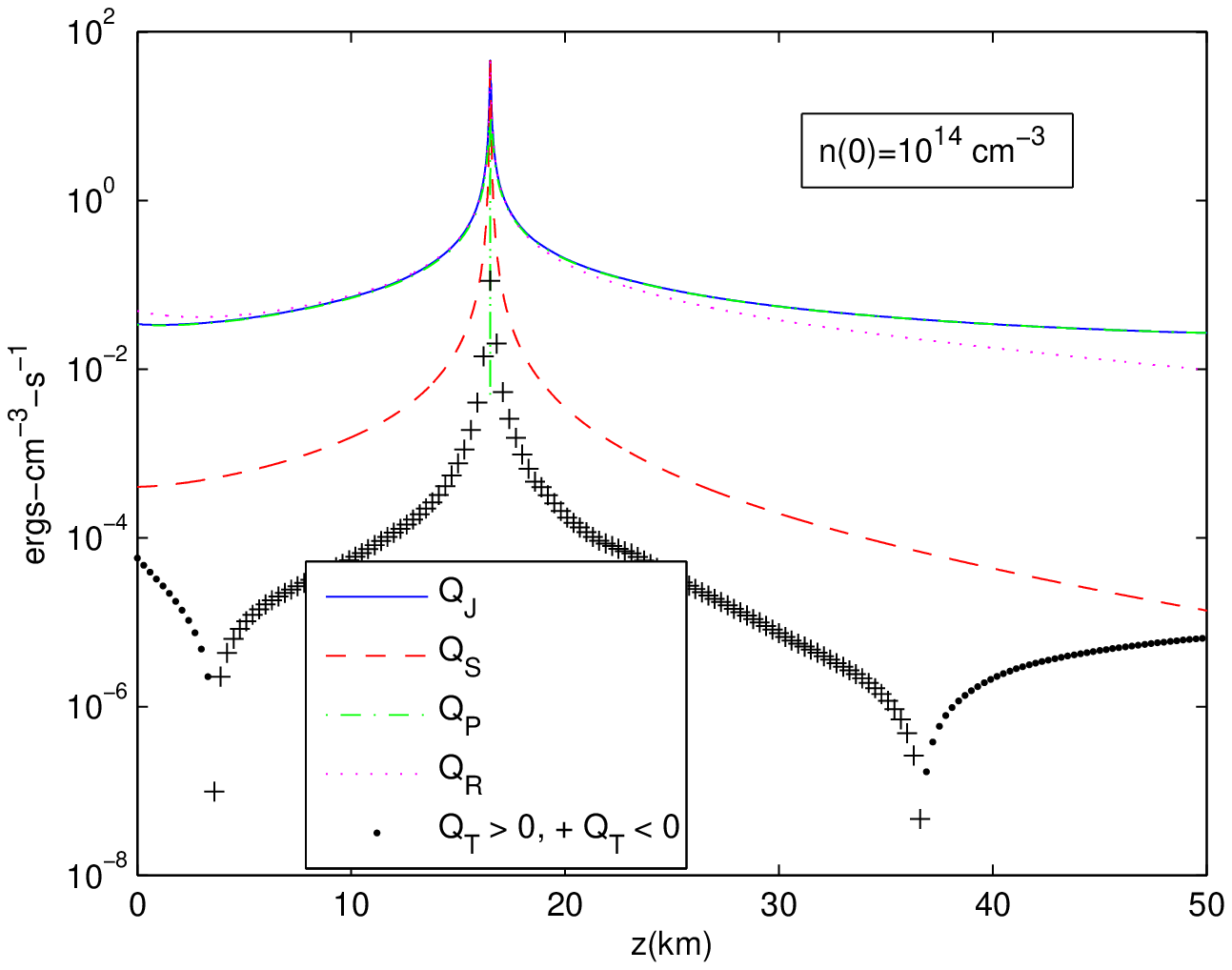}{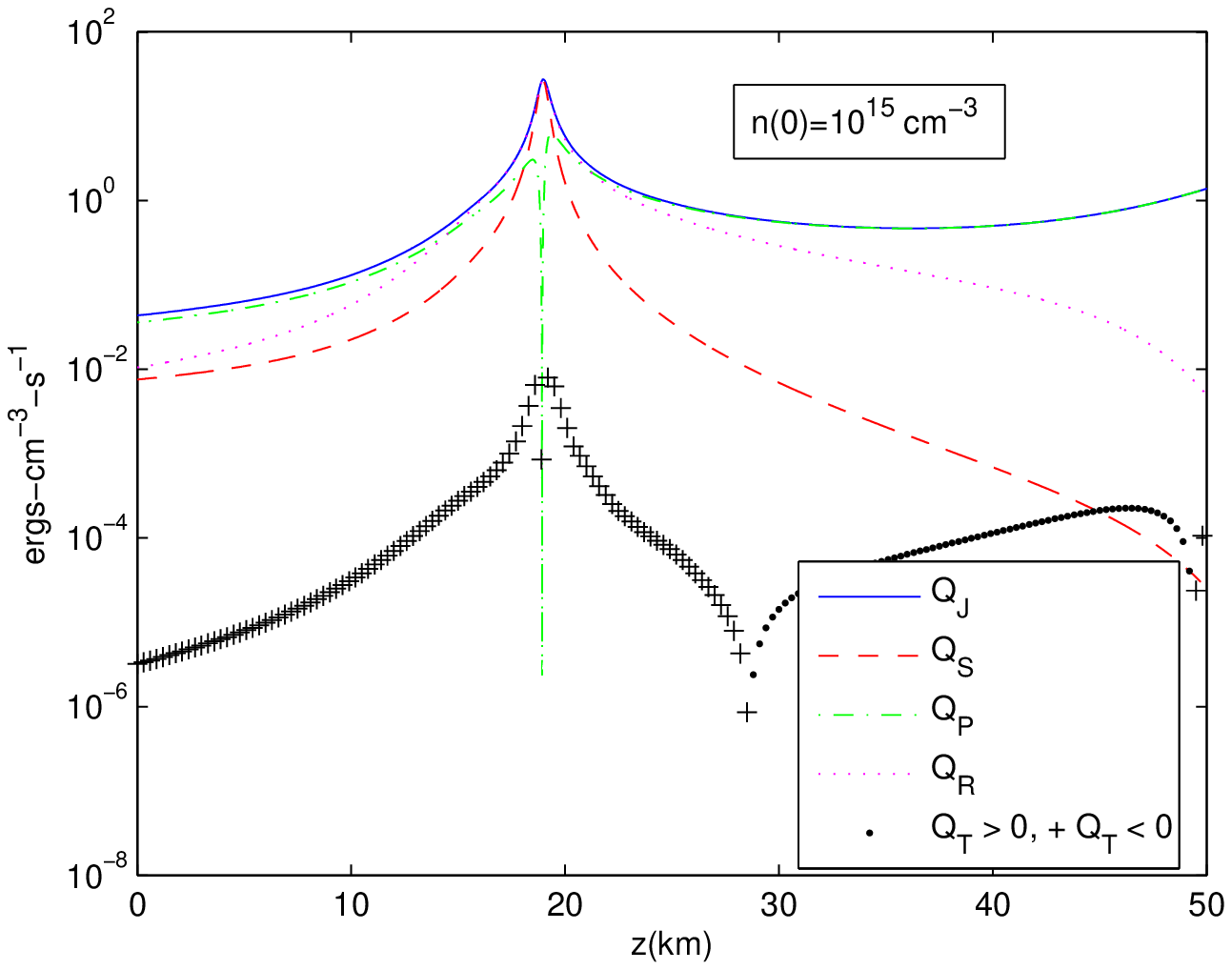}
{Joule heating rate and its components, together with the radiative cooling rate vs. height for the
solution with $n(0)= 10^{14}$ cm$^{-3}$}{and for the solution with $n(0)=10^{15}$ cm$^{-3}$}

\pgjfig{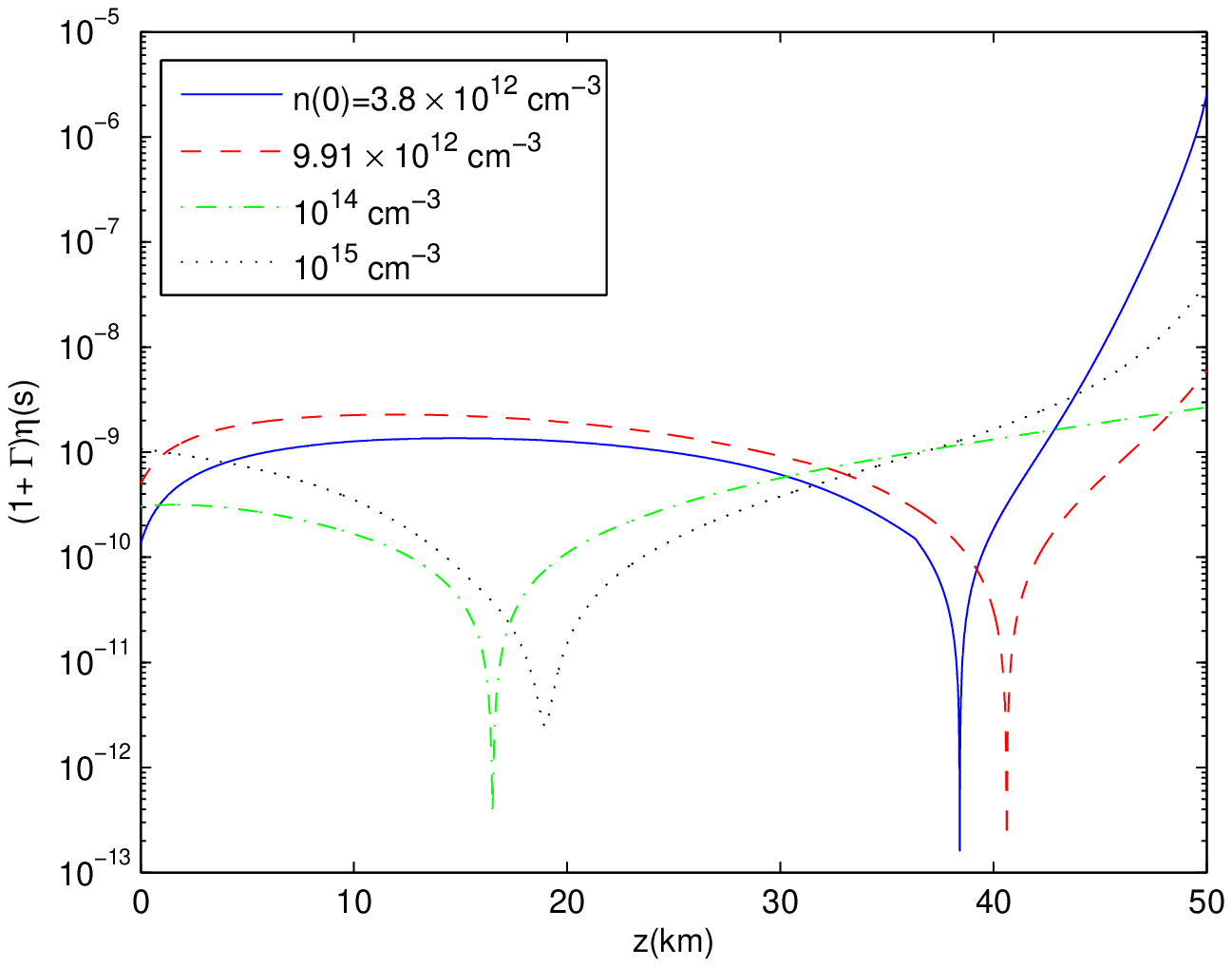}{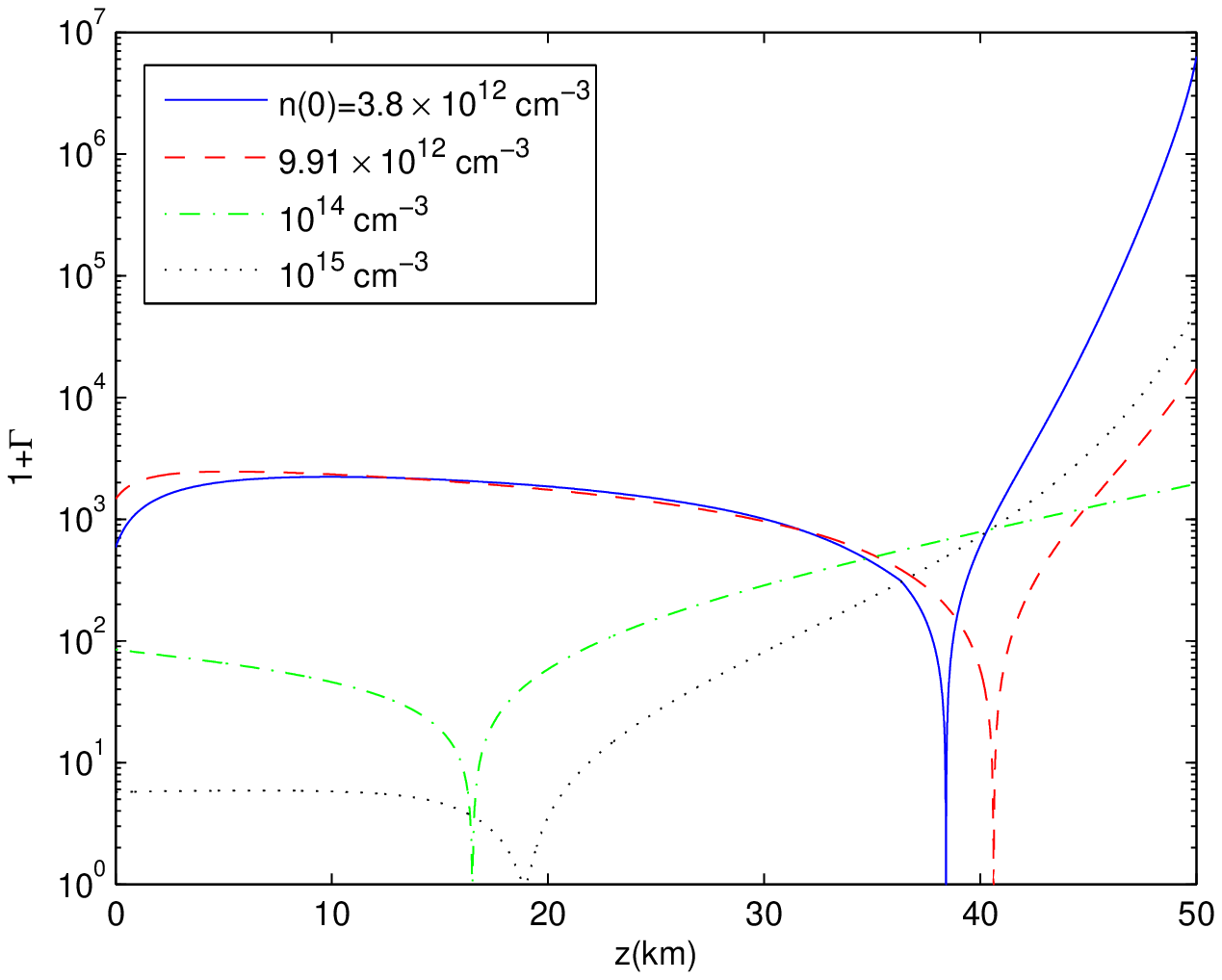}{Total resistivity vs. height}
{and ratio of total resistivity to Spitzer resistivity vs. height. This ratio is the magnetization factor + 1}